\documentclass[english]{WileyChemistry-template}

\usepackage[T1]{fontenc}
\usepackage{libertine}
\usepackage{amsmath}
\usepackage{amssymb}
\usepackage{graphicx}
\usepackage{mathbbol}
\usepackage{color}
\usepackage{indentfirst}
\usepackage{microtype}
\usepackage{etoolbox}
\apptocmd{\sloppy}{\hbadness 10000\relax}{}{}

\usepackage{doi}
\makeatletter
\AtBeginDocument
  {%
    \patchcmd\HyOrg@subequations
      {\alph}{.\arabic}{}{\GenericError{}{Patching failed}{}{}}%
  }
\makeatother
\usepackage{csquotes}
\usepackage{siunitx}
\DeclareSIUnit\C{C}
\usepackage{tabularx}
\newcommand{\SOCn}{\text{SOC}_\text{n}}
\newcommand{\SOCp}{\text{SOC}_\text{p}}
\AtBeginDocument{\colorlet{defaultcolor}{.}}
\makeatletter
\newcolumntype{g}{!{\color{lightgray}\vrule\@width\arrayrulewidth}}
\makeatother

\usepackage{prettyref}
\newrefformat{SI-sec}{the SI Section \ref{#1}}
\newrefformat{SI-eq}{the SI \cref{#1}}
\newrefformat{SI-fig}{the SI Figure \ref{#1}}
\newrefformat{SI-tab}{the SI Table \ref{#1}}
\newrefformat{sec}{Section \nameref{#1}}
\newrefformat{subsec}{Subsection \nameref{#1}}
\newrefformat{eq}{\cref{#1}}
\newrefformat{fig}{Figure \ref{#1}}
\newrefformat{tab}{Table \ref{#1}}

\title{\raggedright Bayesian Parameterization of Continuum Battery Models from Featurized Electrochemical Measurements Considering Noise}

\author{
\begin{minipage}{\textwidth}
    Yannick Kuhn,\textsuperscript{[a]} Hannes Wolf,\textsuperscript{[b]} Arnulf Latz,*\textsuperscript{[c]} Birger Horstmann*\textsuperscript{[c]}
\end{minipage}
}
\newcommand{\affiliation}{
\begin{itemize}
\item[{[a]}] Y. Kuhn\\
German Aerospace Center, Pfaffenwaldring 38-40, 70569 Stuttgart, Germany\\
Helmholtz Institute Ulm, Helmholtzstra{\ss}e 11, 89081 Ulm, Germany\\
E-mail: yannick.kuhn@dlr.de

\item[{[b]}] Dr. H. Wolf\\
BASF SE, Carl-Bosch-Stra{\ss}e 38, 67056 Ludwigshafen am Rhein, Germany

\item[{[c]}] Prof. Dr. A. Latz*, Dr. B. Horstmann*\\
German Aerospace Center, Pfaffenwaldring 38-40, 70569 Stuttgart, Germany\\
Helmholtz Institute Ulm, Helmholtzstra{\ss}e 11, 89081 Ulm, Germany\\
Universität Ulm, Albert-Einstein-Allee 47, 89081 Ulm, Germany\\
E-mail: arnulf.latz@dlr.de\\
birger.horstmann@dlr.de
\end{itemize}
}

\renewcommand{\abstract}{
Physico-chemical continuum battery models are typically parameterized by manual fits, relying on the individual expertise of researchers.
In this article, we introduce a computer algorithm that directly utilizes the experience of battery researchers to extract information from experimental data reproducibly. We extend Bayesian Optimization (BOLFI) with Expectation Propagation (EP) to create a black-box optimizer suited for modular continuum battery models.
Standard approaches compare the experimental data in its raw entirety to the model simulations. By dividing the data into physics-based features, our data-driven approach uses orders of magnitude less simulations.
For validation, we process full-cell GITT measurements to characterize the diffusivities of both electrodes non-destructively.
Our algorithm enables experimentators and theoreticians to investigate, verify, and record their insights. We intend this algorithm to be a tool for the accessible evaluation of experimental databases.
}

\newcommand{\keywords}{
	Electrochemistry \textbullet\ 
	Computational chemistry \textbullet\ 
	Bayesian Optimization \textbullet\ 
    Uncertainty Quantification \textbullet\ 
	Model parameterization
}

\begin{document}
%%%%%%%%%%%%%%%%%%%%%%%%%%%%%%%%%%%%%%%%%%%%%%%%%%%%%%%%%%
%%%%%%%%%%%%%%%%%%%%%%%%%%%%%%%%%%%%%%%%%%%%%%%%%%%%%%%%%%
%%%%%%%%%%%%%%%%%%%%%%%%%%%%%%%%%%%%%%%%%%%%%%%%%%%%%%%%%%

\twocolumn[\vspace{-1.5cm}\maketitle\vspace{-1cm}
	\textit{\dedication}\vspace{0.4cm}]
\small{\begin{shaded}
		\noindent\abstract
	\end{shaded}
}

\begin{figure} [!b]
\begin{minipage}[t]{\columnwidth}{\rule{\columnwidth}{1pt}\footnotesize{\textsf{\affiliation}}}\end{minipage}
\end{figure}

%%%%%%%%%%%%%%%%%%%%%%%%%%%%%%%%%%%%%%%%%%%%%%%%%%%%%%%%%%
%%%%%%%%%%%%%%%%%%%%%%%%%%%%%%%%%%%%%%%%%%%%%%%%%%%%%%%%%%
%%%%%%%%%%%%%%%%%%%%%%%%%%%%%%%%%%%%%%%%%%%%%%%%%%%%%%%%%%

%%%%%%%		 Main Text			%%%%%%% 

\section*{Introduction}

%
%general battery background
%
Batteries are essential for the decarbonization of heavy industry and electricity supply. With their high specific energy density, Li-ion batteries are crucial for efficiently electrifying personal transport, freight transport, and aviation. These applications require materials with optimal energy density, efficiency, and safety. \par
%
%purpose of physics-based battery models
%
Theoretical electrochemists use physics-based continuum battery models \cite{Latz2011} to aid in the search for optimal materials. Physics-based models can predict battery operation and failure modes from material properties rather than artificial fit parameters \cite{Fong2021}. The parameterization of these models is crucial to verify and enhance them. Since the amount of data grows faster than experts can analyze it, such parameterization should be automated. \par
%
%experiments <-> models
%
Parameterization of physics-based battery models may reveal the material properties of a battery from non-destructive measurements. Non-destructive measurements are essential since specific material properties change during the lifetime of a battery. Ageing effects include the formation of a Solid-Electrolyte Interphace (SEI) \cite{Single2016} and Lithium plating \cite{Hein2020}. Tracking these mechanisms is imperative for modelling them. \par
%
%context: non-destructive measuring techniques and what they measure
%
Non-destructive measurements of physical battery parameters usually require special experimental setups. Examples are the Galvanostatic Intermittent Titration Technique (GITT) to measure transport properties \cite{Weppner1977}, Staircase GITT to measure reaction kinetics \cite{Heubner2016}, Electrochemical Impedance Spectroscopy (EIS) to measure the electrode-electrolyte interfacial kinetics \cite{Meddings2020}, or Nuclear Magnetic Resonance (NMR) to measure ionic transport in the electrolyte \cite{Sethurajan2015, Gouverneur2015}. Associated analytical formulas may extract the physical properties of a battery. However, these formulas seldomly adapt well to slight variations in operating conditions, as demonstrated by Horner et al. \cite{Horner2021} in the case of GITT. \par
%
%current state in model parameterization
%
This traditional approach often requires some manual fine-tuning. Recent efforts are devoted to developing automated parameterization methods. Automated algorithms based on physics-based models can cope with less handpicked input. In this case, the task is to match the simulated model to the experimental data. Usually, this will involve repeated model evaluations by an optimization algorithm. Commonly used optimizers for battery models are Least Squares \cite{Bizeray2019}, Monte Carlo \cite{Bolay2021}, and Genetic Algorithms \cite{Forman2012, Li2022}. All of these have a stochastical element for efficiency but do not give error bars for their parameterizations. Bayesian algorithms such as Kalman Filters \cite{Plett2006, Plett2004} and Markov-Chain Monte Carlo \cite{Aitio2020} simultaneously provide the optimal parameters and their uncertainty. Another method of adding error bars is to repeat the parameterization with artificial noise in the data that mimics the original noise \cite{Sethurajan2019}. For best results, multiple approaches to parameterization have to be combined \cite{Xu2022}. Model substitution is a technique primarily employed by data scientists to accelerate parameterization. There, a Neural Network or other stochastical classifier gets fitted to a physics-based model and is then used instead in an optimizer \cite{Yamanaka2020, Bills2020, Houchins2020, Li2021}. Imaging techniques also give essential input parameters for spatially resolved simulations and have to concern themselves with the uncertainty propagation from imaging to simulation outcome \cite{Zhao2020, Krygier2021, Biebl2019, Furat2021, Scharf2021}. The State-of-Health estimation of degradation modes may also benefit from a physical understanding of the degradation processes involved \cite{Birkl2017}. \par
%
%current direction in model parameterization: Bayesian
%
Bayesian algorithms directly match model simulations to experimental data and to the uncertainty in that data. The considerable noise in battery characterization measurements is thus well handled in Bayesian algorithms. To the best of our knowledge, only two distinct types of Bayesian algorithms are in use for the simultaneous estimation of parameters of continuum battery models and their uncertainty. On the one hand, there are Dual Kalman Filters (DKF) \cite{Plett2006, Plett2004}, which estimate the state of the simulated battery over time and can also optimize the model parameters. The high flexibility of DKF estimations has a drawback: DKFs have to be tuned perfectly and require an increased effort for integration with the model simulation. Otherwise, their results are technically correct but physically unreasonable. On the other hand, there are Markov-Chain Monte Carlo (MCMC) algorithms \cite{Aitio2020}, which run tens of thousands of simulations to find the one that fits the experiment the best. Stability is thus enhanced at the cost of run-time, so only simplified battery models that simulate in milliseconds are currently used \cite{Aitio2020, Bizeray2019}. \par
%
%parameterization challenges: identifiability, oversimplification
%
Uncertainty Quantification (UQ), i.e., estimation of the precision of parameter fits, is a significant advantage of Bayesian algorithms. Predicting battery performance and failure modes relies heavily on the accuracy of the model parameters. However, the precision with which a parameter is identifiable in a given measurement gets often neglected \cite{Tang2019, Chen2020, Maheshwari2016}. In other words, it is then doubtful that the measurement uniquely identifies the fit parameters. \textcolor{defaultcolor}{For example, Escalante et al. \cite{Escalante2021} apply Bayesian inference to charge-discharge cycles and find that they are barely suitable for battery model parameterization. Especially in the presented case of a partial Differential-Algebraic Equation model, parameterizing it is an \textquote{inverse problem}. \textquote{Inverse problems} are ill-posed in that they admit a family of infinitely many approximate solutions rather than one unambiguous one, as \textquote{direct problems} do. With Bayesian algorithms, these many solutions can be ranked systematically by their assigned probabilities and parameter interdependencies can be formally analyzed via their correlations. Still, it is important to keep the number of estimated model parameters small.} Systematic model parameterizations \cite{Bizeray2019, Park2018, Zhao2020a, Forman2012} are crucial for the reproducibility and transferability of the results. Further theoretical and experimental developments may build upon such reusable prior work. \par
%
%our method and results
%
This article introduces a new Bayesian algorithm, called EP-BOLFI, for the automated estimation of model parameters by applying it to continuum battery models. Our algorithm requires one order of magnitude fewer simulations \textcolor{defaultcolor}{and is more stable} than the MCMC approach, requiring only a consumer-grade computer. Our algorithm is independent of the battery model and the experiment, so different battery models can be examined quickly for identifiability from any measurement. \textcolor{defaultcolor}{Battery modelling experts can input their expertise to improve the parameterization by segmenting the data into \textquote{features}. Other than taking segments of the raw data, using the parameters of fitting functions on the data as features can improve weighting of the information contained in the data. While we deliberately do not automate the feature selection, it is possible to automatically select from a set of candidate fitting functions \cite{Gasper2021}.} \par
%
%paragraph of contents
%
We briefly show the battery model's equations we apply our algorithm to as an example in \prettyref{subsec:Physics}. We give a short introduction to the Bayesian idea in general in \prettyref{subsec:Bayes}. A deeper understanding of the two Bayesian algorithms we use follows, namely Expectation Propagation \cite{Barthelme2015} in \prettyref{subsec:EP} and a specific Bayesian Optimization implementation \cite{Gutmann2016} in \prettyref{subsec:BOLFI}. We compare our algorithm with the state of the art in Bayesian parameterization, namely MCMC, in \prettyref{sec:performance}. The properties of the lithium-ion battery we measured, the selection of the unknown parameters we fit and the setup of the battery model we fit are in \prettyref{sec:Experimental}. We then show the application of these algorithms to full-cell GITT measurements in \prettyref{sec:Results}. A discussion of the results of our algorithm follows in \prettyref{sec:Discussions}. We conclude this paper with our findings in \prettyref{sec:Conclusion}. \par

\section*{Theory} \label{sec:Theory}

\subsection*{Physics-Based Battery Models} \label{subsec:Physics}

\begin{figure}[t]
    \includegraphics[width=0.99\columnwidth]{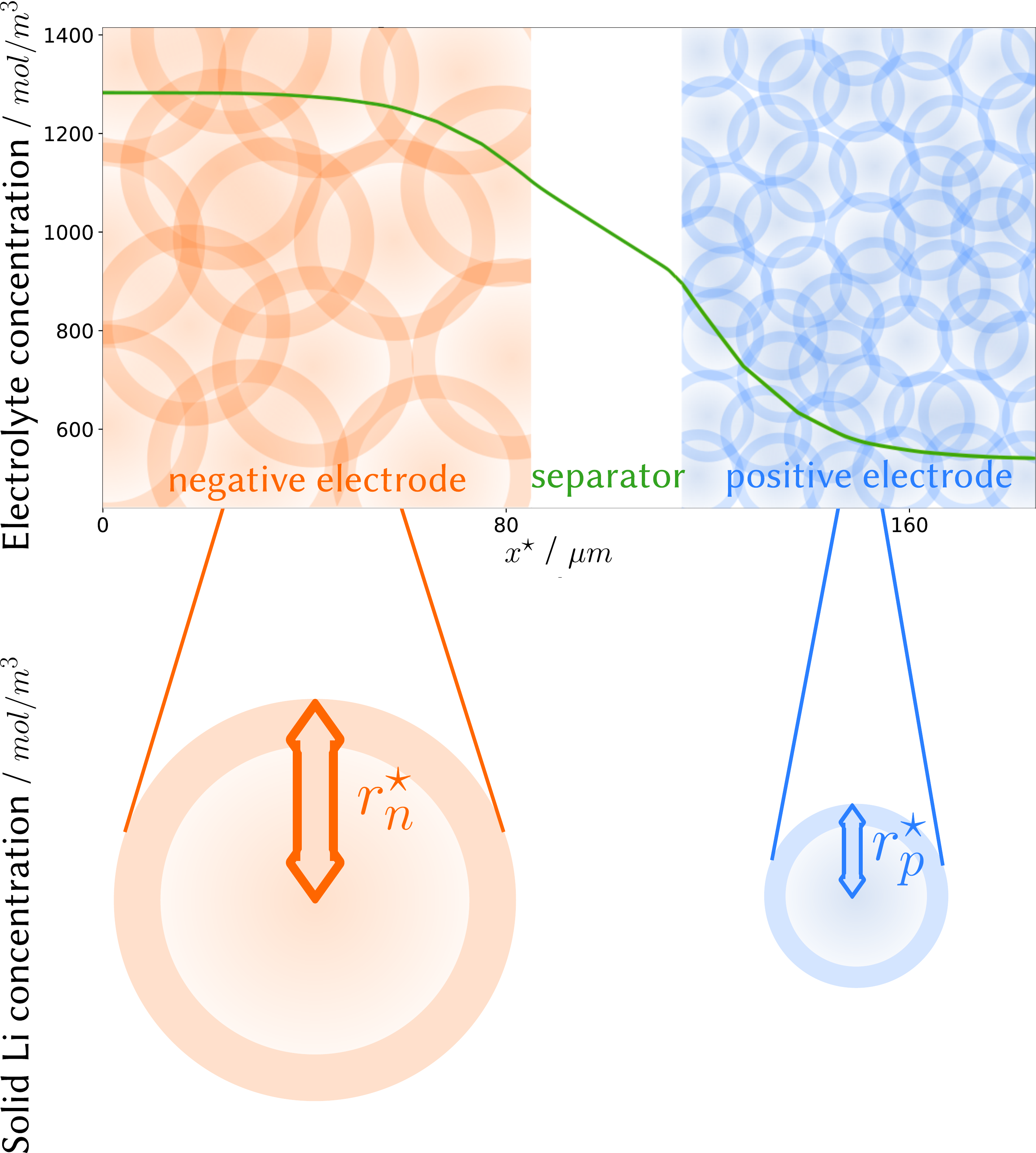}
    \caption{Schema of a battery cell as represented by the DFN model in \prettyref{eq:nodim}. All particles at the same place in \(x\)-direction are averaged into one representative particle.}
    \label{fig:DFN}
\end{figure}

\begin{table*}[t]
\begin{center}
\caption{1D+1D physics-based continuum battery model equations}
\label{eq:nodim}
\begin{subequations}
\begin{align}
    &\text{electrolyte cation \textcolor{defaultcolor}{flux}} &i_{e,k} &= \varepsilon_k^{\beta_k} \hat{\kappa}_e \kappa_e \left(-\partial_x\phi_{e,k} + 2 (1 - t_+) (1 + \partial_{ln(c_e^*)}\ln(f_+)) \partial_x\ln(c_{e,k})\right) \label{eq:nodim-ie}&& \\
    &\text{electrolyte cation source} &\partial_xi_{e,k} &= i_{se,k}\text{ with } i_{se,s} := 0 \label{eq:nodim-partialie}&& \\
    &\text{electrolyte cation molar \textcolor{defaultcolor}{flux}} &N_{e,k} &= -\varepsilon_k^{\beta_k} D_e(c_{e,k}) \partial_xc_{e,k} + \frac{C_e t_+}{\gamma_e} i_{e,k} \label{eq:nodim-Ne}&& \\
    &\text{electrolyte cation diffusion} &\partial_tc_{e,k} &= -\frac{1}{\varepsilon_k} \partial_x\frac{N_{e,k}}{C_e} + \frac{1}{\gamma_e\varepsilon_k} \partial_xi_{e,k} \label{eq:nodim-ce}&& \\
    &\text{electrode electronic \textcolor{defaultcolor}{flux}} &i_{s,k} &= -\sigma_k \partial_x\phi_{s,k},\label{eq:nodim-is}&& \\
    &\text{electrode electronic source} &\partial_x i_{s,k} &= -i_{se,k} \label{eq:nodim-partialis}&& \\
    &\text{electrode lithium diffusion} &\partial_t c_{s,k} &= -\frac{1}{r_k^2} \partial_{r_k} \left( -r_k^2 \frac{D_{s,k}(c_{s,k})}{C_k} \partial_{r_k} c_{s,k} \right) \label{eq:nodim-cs}&& \\
    &\text{interface reaction current density} &i_{se,k} &= \frac{\gamma_k}{C_{r,k}} i_{se,k,0} \left( e^{+\alpha_{n,k} z_k \eta_k} - e^{-\alpha_{p,k} z_k \eta_k} \right) \label{eq:nodim-ise}&& \\
    &\text{interface overpotential} &\eta_k &= \phi_{s,k} - \phi_{e,k} - U_k\left( c_{s,k}\Big|_{r_k=1} \right) - \frac{\ln(c_{e,k})}{z_k} \label{eq:nodim-eta}&&
\end{align}
\end{subequations}
\hrule
\end{center}
\end{table*}

%nomenclature of physics-based models
Physics-based continuum battery models consist of partial Dif-\\ferential-Algebraic Equations (DAE). Solving partial DAEs on a microstructure-resolving 3D grid requires high-performance computers. Volume-averaged 3D model simulations still take a couple of days \cite{Schmitt2020}. Hence, volume-averaged 1D+1D models \cite{Latz2011} have been developed as a suitable compromise between accuracy and speed. \par
%why we do not go lower than the DFN
In this paper we will use the Doyle-Fuller-Newman (DFN) 1D+1D model. There are further simplifications down to 1D, especially the Single Particle Model (SPM) and the SPM with electrolyte modification (SPMe) \cite{Marquis2019}. However, we found that these cannot accurately describe the GITT experiments which represent the main example in this article. \par
%what does a 1D+1D model model
1D+1D models \cite{Latz2011} represent a battery on the cell level: porous anode, porous separator, and porous cathode. Inside this porous structure, the ionically conducting electrolyte is modelled as a continuum. The electrodes are simplified to a cluster of spherical, homogenous particles in which intercalated lithium diffuses. These representative particles are separated by electrolyte and not in direct contact. \par
%why 1D+1D is the lowest acceptable simplification
1D+1D models are still too costly to evaluate for parameter identification schemes that require hundreds of thousands of simulations. For example, Aitio et al. \cite{Aitio2020} instead use an asymptotically simplified model, the Single-Particle Model with electrolyte (SPMe) \cite{Marquis2019}. The SPMe consists of ordinary differential equations (ODE) only. These ODEs simulate in milliseconds, but the SPMe gives inaccurate results at currents over \(\SI{1}{\C}\). \par
%DFN model summary
In \prettyref{eq:nodim} we summarize the common ancestor of all 1D+1D models, the Doyle-Fuller-Newman (DFN) model \cite{Doyle1993}, in a non-dimensionalized form laid out by Marquis et al. \cite{Marquis2019}. \prettyref{fig:DFN} visualizes the level of detail the DFN describes: all material properties and dynamics are assumed to be homogenous perpendicular to the line between the current collectors. Along this line, the DFN describes lithium(-ion) concentrations and potentials in the electrodes and the electrolyte. \textcolor{defaultcolor}{The DFN model parameters are summarized in \prettyref{tab:parameters} and \prettyref{tab:scalings}. The boundary conditions and initial conditions that complete the DAE system are stated in \prettyref{SI-sec:boundary_conditions}.} \par

\subsection*{Introduction to the Bayesian Principle (for Likelihood-Free Inference)} \label{subsec:Bayes}

\begin{figure}[t]
    \includegraphics[width=\columnwidth]{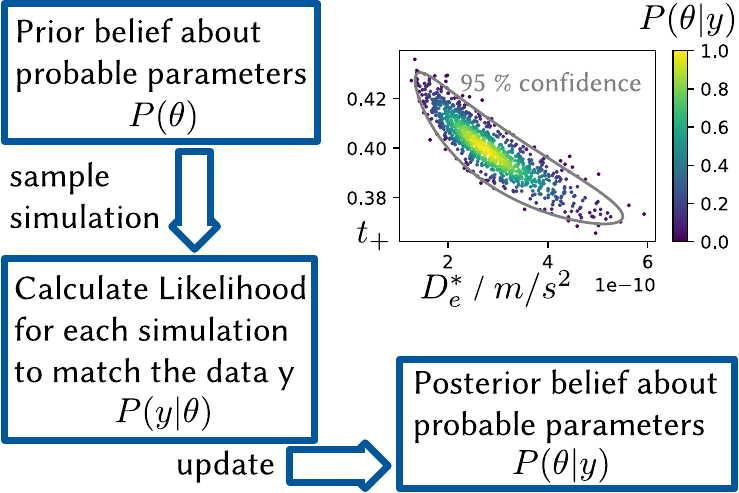}
    \caption{Monte Carlo algorithm for Bayes' Theorem. \(y\) is a placeholder for \textquote{data} and \(\theta\) is a placeholder for \textquote{model parameters}.}
    \label{fig:BayesTheorem}
\end{figure}

%A general introduction to Bayesian methods is given by van Oijen in his book \textquote{Bayesian Compendium} \cite{VanOijen2020}. Example applications of Bayesian methods are given by Held and Bové in their book \textquote{Likelihood and Bayesian inference and computation} \cite{Held2010}. Implementation of, amongst other things, Bayesian methods are given by Subair and Thron in their book \textquote{Implementations and Applications of Machine Learning} \cite{Subair2020}.
%why Bayesian algorithms
Updating prior knowledge given new evidence: this is the central idea behind any Bayesian algorithm. The preconditioning of estimation problems with prior knowledge enables Bayesian algorithms to require far fewer data points than empirical approaches. Some standard textbooks for Bayesian inference are References \citenum{VanOijen2020, Held2010, Subair2020}. \par
%what does a Bayesian algorithm do
\textquote{Probability} in the Bayesian context describes the level of uncertainty in the data rather than randomness. The result of any Bayesian algorithm is a probability distribution providing a range of estimates with their respective \textquote{probabilities} to be the correct estimate. The estimated probability distribution also includes information about how the estimated model parameters influence the measurement individually and collectively \cite{VanOijen2020, Held2010, Subair2020}. \par
%how does a Bayesian algorithm work
Here we describe the basic Bayesian algorithm, which we outline in \prettyref{fig:BayesTheorem}. Firstly, transform the prior knowledge about the range of probable model parameters into a probability distribution \(P(\text{parameter})\), called Prior or \textquote{belief}. Secondly, take random samples from the Prior and calculate the so-called Likelihood \(P(\text{data}|\text{parameter})\) for each parameter sample. The Likelihood represents the \textquote{Likelihood} for each model simulation to match the experimental data. A simple approximation for the Likelihood is a Gaussian distribution with the simulation-experiment agreement as its expectation value and some variance \cite{Aitio2020}. Finally, multiply the Likelihood with the Prior and normalize the result into a probability distribution \(P(\text{parameter}|\text{data})\), called Posterior. The third step is \prettyref{eq:BayesTheorem}, called Bayes' Theorem. Bayes' Theorem ensures that the Posterior reasonably updates the prior belief about probable parameter candidates \cite{VanOijen2020, Held2010, Subair2020},
\begin{equation}
    \label{eq:BayesTheorem}
    P(\text{parameter}|\text{data}) \propto P(\text{data}|\text{parameter}) \cdot P(\text{parameter}).
\end{equation}
%
%short-hand
In the following, we abbreviate \textquote{parameter} as \(\boldsymbol{\theta}\) and \textquote{data} as \(y\), as shown in \prettyref{fig:BayesTheorem}. Bayes' Theorem then reads as \(P(\boldsymbol{\theta}|y) \propto P(y|\boldsymbol{\theta}) \cdot P(\boldsymbol{\theta})\). \par
%why are classical Bayesian algorithms not applicable here
The performance of a Bayesian algorithm crucially depends on the quality of the Likelihood approximation. A rough approximation of a Likelihood with a Gaussian \cite{Aitio2020} might lead to slow convergence or wrong estimates. And in the case of 1D+1D models without analytic solutions, a better Likelihood cannot be derived. However, the approximation that the Posterior is Gaussian is usually justified if the estimated parameters are identifiable from the data, given the Central Limit Theorem \cite{VanOijen2020, Held2010, Subair2020}. \par
%what is LFI
Likelihood-Free Inference (LFI), also called Approximate Bayes-\\ian Computation (ABC), is a class of algorithms that approximate the unknown Likelihood from model evaluations. The shape of the Likelihood in LFI algorithms results from a Machine Learning procedure on simulated data samples, e.g., Cross-Validation \cite{Stone1974}. Instead of a Likelihood, LFI algorithms only need to know the discrepancy measure between a simulated measurement and an experiment. The added flexibility enables one to try out different models freely \cite{Sunnaker2013}. \par
%what does everyone else do
In contrast to DKFs, the discrepancy measure in LFI algorithms may encompass the whole experimental data rather than only one point in a time series at a time. The difference to MCMC methods such as the Metropolis-Hastings algorithm used by Aitio et al. \cite{Aitio2020} is that the quality of parameter guesses is judged by the approximated Likelihood rather than by the Posterior. Using the Posterior like Metropolis-Hastings may result in confirmation bias since it mixes the Prior into evaluating the quality of parameter fits. \par
%announcing the next two sections
The following sections present the two LFI algorithms we employ to minimize the required number of model simulations. Despite the improved generality and accuracy of our approach, we achieve a reduction rather than an increase in computational load compared to Metropolis-Hastings. \par

\subsection*{Expectation Propagation} \label{subsec:EP}

%what is EP
Expectation Propagation (EP) was developed by Minka in his PhD thesis \cite{Minka2001} and later revisited in a paper \cite{Minka2013}. EP has uses for training neural networks \cite{Courbariaux2015}, object detection \cite{Nam2014}, speech recognition \cite{Dahl2010}, and signal processing \cite{Loeliger2007}. We present Barthelmé et al.'s adaptation of EP to Likelihood-Free Inference (LFI) \cite{Barthelme2015} using BOLFI as LFI algorithm (see \prettyref{subsec:BOLFI} for BOLFI). For further reading, we recommend the general introduction to EP in \cite{Bilmes1998}, Chapter 3.6. \par
%why is EP
The great advantage of EP for LFI is splitting the data into multiple segments or discrete features. These features give the sampling algorithm within EP the more straightforward task of matching one part of the data at a time. \par
%what does EP do: moment-matching
EP is proven to work accurately and efficiently based on its two core principles. First, the moments of the approximated Posterior converge to the moments of the actual Posterior. Examples of moments for probability distributions are the expectation value and the variance. Formally, moment-matching is only guaranteed if Posteriors get selected from an exponential family. Further details are summarized in \prettyref{SI-sec:expfam}. \par
%what does EP do: information-propagation
Second, EP efficiently optimizes multiple measurement parts together by iterating through them. From now on, we call these measurement parts \textquote{features} and denote them with \(y_i\). EP searches the optimal parameter set for each feature in the range of probable candidates for the other features. Examples of possible features range from measurements in time to characteristic values, like decay rates. \par
%how does one use EP
The user sets up EP with the following two inputs \cite{Barthelme2015}. One is a functional definition of the features in simulation and data, which defines the cost function. The other is a Prior \(P_0\), which represents the initial belief \(P_0(\boldsymbol{\theta})\) for every vector of parameter values \(\boldsymbol{\theta}\) and defines the \textquote{initial value} of the Bayesian algorithm. Additionally, one may pre-define the Likelihood sites \(p_i\) for each of the \(i=1,...,n\) features, but for initialization \textquote{uninformative Priors} \(p_i(\boldsymbol{\theta}) = 1\) are appropriate. The EP algorithm then expresses the Posterior as the product of the Prior and each Likelihood site, as shown in \prettyref{eq:factorization}.
\begin{equation}
    P \propto \left( \prod_{i=1}^n p_i \right) \cdot P_0.\label{eq:factorization}
\end{equation}

\begin{figure}[t]
    \includegraphics[width=\columnwidth]{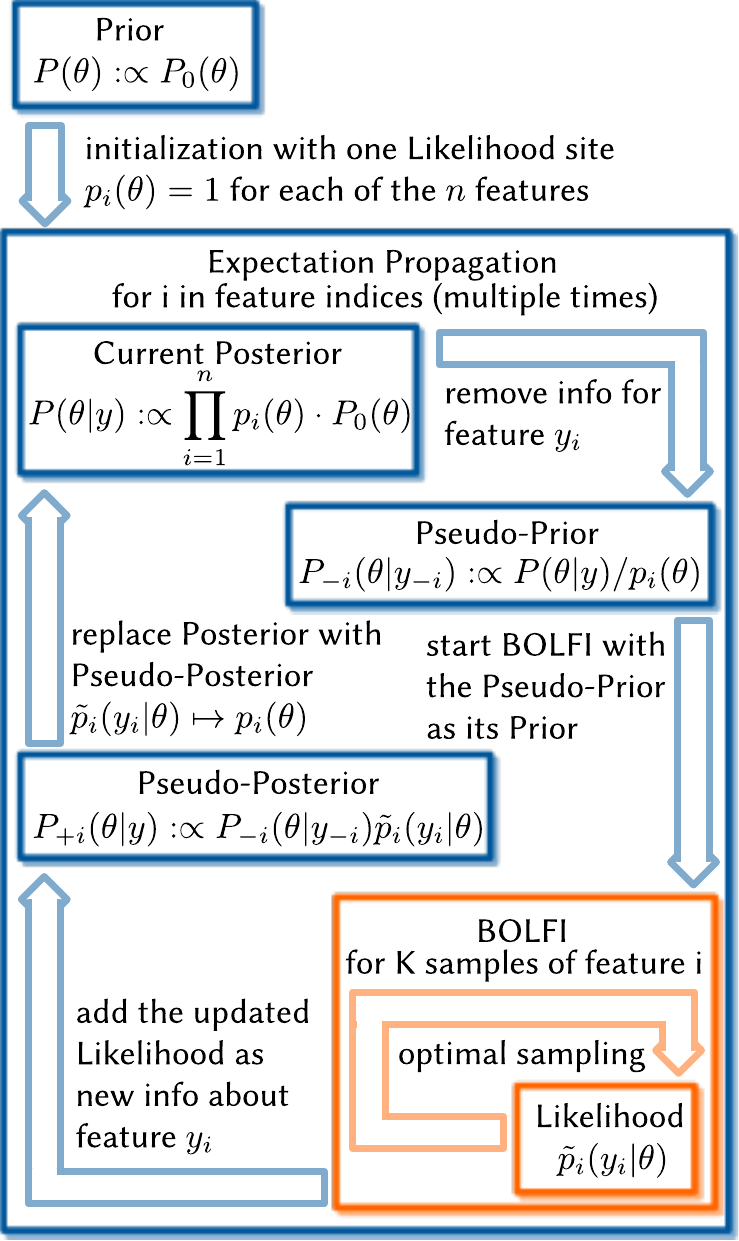}
    \caption{Iterative Expectation Propagation (EP) algorithm. Choose a random feature \(y_i\), create a search space \(P_i\) by omitting the factor for \(y_i\) in \(P(\theta|y)\), sample that search space to obtain a local Posterior \(\tilde{p}_i\) and replace \(p_i\) with it. BOLFI is presented in \prettyref{subsec:BOLFI} and is our choice for the LFI algorithm inside EP.}
    \label{fig:EP_Loop}
\end{figure}

%how does EP work: paraphrased
The procedure for EP \cite{Barthelme2015} factorizes the simulation and data into individual features. EP then iterates through all features multiple times. For each feature, BOLFI optimizes the model parameters to fit this feature. The result of BOLFI is a more precise probability distribution of the model parameters. EP utilizes this result to update its distribution of probable model parameters for all features. EP uses the new Posterior as input for next feature optimization. \par
%how does EP work: algorithm
The algorithm of EP \cite{Barthelme2015} iterates through the features \(y_i\) of the whole data \(y\) until a stop criterion is met, as shown in \prettyref{fig:EP_Loop}. After initialization of the current Posterior with the Prior and the initial Likelihood sites, there are four steps in each iteration. Firstly, produce a \textquote{Pseudo-Prior} \(P_{-i} :\propto p / p_i\) by omitting the corresponding Likelihood site \(p_i\) from the current Posterior. Secondly, use an LFI algorithm like BOLFI to compute an approximation \(\tilde{p}_i\) to the Likelihood of the selected feature. We introduce BOLFI in \prettyref{subsec:BOLFI}. Thirdly, produce a \textquote{Pseudo-Posterior} \(P_{+i} :\propto P \tilde{p}_i / p_i\) by replacing \(p_i\) with the updated \(\tilde{p}_{i}\). BOLFI basically takes the Pseudo-Prior as its Prior, approximates a Likelihood, and outputs the Pseudo-Posterior as its Posterior, i.e., the product of Pseudo-Prior and approximated Likelihood. Finally, replace the current Posterior with the Pseudo-Posterior for the next iteration and start the next iteration. \par
%disadvantage of EP
\textcolor{defaultcolor}{There is one caveat inherent in the capability to run multiple loops through all features. At the step where \(\tilde{p}_i\) replaces \(p_i\), more precisely a projection of \(\tilde{p}_i\) replaces \(p_i\). The projection to a certain type of probability distribution ensures that the replacement results in a Pseudo-Posterior that retains its type of probability distribution. In our case, all involved quantities \(P_0\), \(p_i\) and \(\tilde{p}_i\), and hence, \(P\), are normal distributions. Any multi-modality in \(\tilde{p}_i\) before projection gets lost and only results in a wider normal distribution after projection that smooths over the multiple modes.}
%EP dampening
In the algorithm description above, we omit dampening for clarity. With a dampening parameter \({\alpha\in]0,1[}\), dampening is introduced to EP by linearly interpolating between \(P(\theta|y)\) prior to each site update and \(P_{+i}(\theta|y)\) in terms of their so-called \textquote{natural parameters}. \cite{Barthelme2015} We show the detailed formula in \prettyref{SI-sec:EP}. \par
%EP benefits: sample efficiency
The advantage of subdividing the discrepancy between experiment and simulation into features is the reduced number of simulations needed for convergence. Any LFI algorithm must decide which of the sample simulations it should include in the Likelihood approximation. The more complex the Likelihood is, the more samples the LFI algorithm ultimately discards due to the so-called \textquote{curse of dimensionality}. With each additional \textquote{dimension of complexity}, the computational effort to deal with it grows exponentially. \textcolor{defaultcolor}{By \textquote{dimension of complexity}, we refer to the effective dimension of the information in the data. For example, suppose a set of fitting functions describe the data up to noise. Then, the information contained has a dimension not larger than the number of fit parameters rather than the number of data points.} Using a subset of the whole discrepancy with a lower dimension gives a much higher chance of any random sample being close to the optimum of this subset \cite{Barthelme2015}. \par
%EP benefits: continuous preconditioning
The sequential handling of the features further reduces the computational complexity. Each fitted feature, i.e., updated \(p_i\), preconditions the estimation task much like the Prior does at the beginning of every Bayesian algorithm. Hence, EP takes no unnecessary samples that contradict an already fitted feature. This preconditioning is most efficient when the features are uncorrelated, which gives an upper limit on the sensible number of features. \par
%EP convergence
Minka \cite{Minka2001} has proven that EP converges in the sense of minimizing the so-called Kullback-Leibler Divergence (KLD). KLD is not an actual distance metric for distributions, but its second derivative gives a pragmatic approximation of the usually used Fisher Information Metric (FIM) \cite{Kullback1951}. EP converges on the mean of the posterior with quadratic speed \(O(1/K^2)\) when the posterior tends to a normal distribution, as stated in Barthelmé et al. \cite{Barthelme2015}. Here, \(K\) is the total number of simulation samples. \par
%%
%\begin{equation}
%    KL(\pi||P) := \int_\Theta\ \pi(\boldsymbol{\theta}) \ln\left( \frac{\pi(\boldsymbol{\theta})}{P(\boldsymbol{\theta})} \right)\ d\boldsymbol{\theta}.\label{KLdivergence}
%\end{equation}
%%
%Here, \(\pi\) and \(P\) are some probability distributions over the same domain. \par

\subsection*{Bayesian Optimization (for Likelihood-Free Inference)} \label{subsec:BOLFI}

\begin{figure}[t]
    \includegraphics[width=\columnwidth]{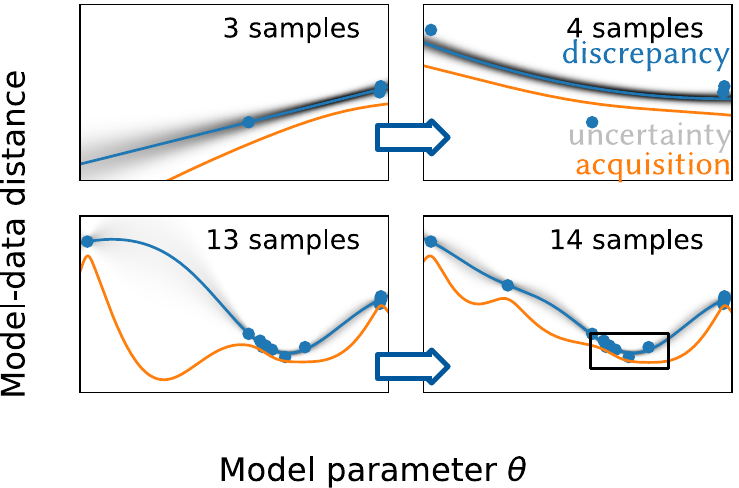}
    \caption{BOLFI sampling paradigm. \textcolor{defaultcolor}{A Gaussian Process approximates the relationship between the model-data distance and the model parameters, labelled as \textquote{discrepancy} with \textquote{uncertainty}. An \textquote{acquisition} function can be calculated from this Gaussian Process. The minimum of the \textquote{acquisition} function gives the most informative next sample.} Circles indicate the sample points. The black border in the bottom right plot indicates the cutout presented in \prettyref{fig:Likelihood}a.}
    \label{fig:BOLFI}
\end{figure}

%what is BOLFI
Bayesian Optimization for Likelihood-Free Inference (BOLFI) was developed by Gutmann et al. \cite{Gutmann2016}. Applications of their algorithm include the fields of cosmology \cite{Alsing2019}, ecology \cite{Fer2018}, genetics \cite{Arnold2017}, and neurobiology \cite{Kangasraasio2017}. We use an implementation of BOLFI by Lintusaari et al. in the software ELFI \cite{Lintusaari2018}. For further reading, we recommend the introduction to BOLFI in \cite{Lintusaari2017}. To the best of our knowledge, we are the first to combine BOLFI with Expectation Propagation. \par
%BOLFI benefits
Bayesian Optimization takes the idea of preconditioning an estimation task with a Prior one step further: this class of algorithms utilizes each sample to optimize the choice of the following sample. This short-circuiting of the usual Bayesian recipe allows for the active selection of the most informative following sample, where standard Bayesian algorithms draw samples randomly. \prettyref{fig:BOLFI} visualizes how BOLFI \textquote{explores} the parameter search space before \textquote{exploiting} the region where BOLFI expects the optimal parameter set. A sample is chosen by minimizing its \textquote{acquisition} function in \prettyref{eq:acquisition}. To this aim, a Gaussian Process is trained on the preceding simulation samples and acts as a fast surrogate for the intractable discrepancy-parameter relationship to enable this minimization. \par
%why BOLFI with EP
BOLFI works best for low sample numbers. This is because it involves the inversion of a matrix with rank equal to the sample size at each sample. Expectation Propagation helps in this respect as it keeps the sample size small. Each feature update resets the samples to 0 and provides a more precise Prior for the next iteration. While this resolves the most significant weakness of BOLFI, it requires that the form of the current Posterior does not change over multiple site updates. Here, we limit ourselves to Gaussian distributions. The transformations can limit the parameters to positive numbers by taking the logarithm or to an interval by taking the tangent. Please note that any proper probability distribution can be optimized with BOLFI. \par
%BOLFI algorithm: initializing
For each feature with index \(i\), the algorithm of BOLFI \cite{Gutmann2016} starts with a user-defined number \(K_0\) of (quasi-)random samples from the Prior. To achieve optimal integration efficiency, these samples stem from a Sobol sequence \cite{Sobol1967}. These samples \(\boldsymbol{\theta}_k\) and the deviations between simulation and experiment \(\log(\|y_i(\boldsymbol{\theta}_k) - y_i^\star\|)\) constitute the data that BOLFI trains a surrogate function on. Here, \(y_i^\star\) is the current feature of the experimental data and \(y_i(\boldsymbol{\theta}_k)\) is the simulated feature for the parameter set \(\boldsymbol{\theta}_k\). For illustration, \prettyref{fig:BOLFI} shows as blue dots \(\log(\|y_i(\boldsymbol{\theta}_k) - y_i^\star\|)\) over \(\boldsymbol{\theta}_k\), where the blue line labelled \textquote{discrepancy} would be that surrogate function. \par
The training points \(\log(\|y_i(\boldsymbol{\theta}_k) - y_i^\star\|)\) are assumed to follow a Gaussian Process with parabolic mean function. From this model, BOLFI applies a filter to this Gaussian Process to calculate the approximation to the model-simulation discrepancy \({\log(\|y_i(\cdot) - y_i^\star\|)}\),
\begin{equation}
\label{eq:surrogate}
\log(\|y_i(\boldsymbol{\theta})-y_i^\star\|) \sim \mathcal{N}(\mu_K(\boldsymbol{\theta}), v_K(\boldsymbol{\theta}) + \sigma^2),
\end{equation}
where \(\sigma^2 > 0\) is adjusted automatically. We show the derivation of this Posterior in \prettyref{SI-sec:BOLFI}. \par
%BOLFI algorithm: sampling optimization
\(\mu_K\) and \(v_k\) constitute a differentiable surrogate function for the intractable model-simulation discrepancy function\\\({\log(\|y_i(\cdot)-y_i^\star\|)}\). Following the initial regression to the \(K_0\) Sobol quasi-random samples, further samples are generated from the surrogate in \prettyref{eq:surrogate} in the following manner, called \textquote{lower confidence bound selection criterion} \cite{Gutmann2016}. The next sample gets taken as the minimum value of the so-called acquisition function
\begin{equation}
    \mathcal{A}_K(\boldsymbol{\theta}) := \mu_K(\boldsymbol{\theta}) - \sqrt{\eta_K^2 v_K(\boldsymbol{\theta})}.\label{eq:acquisition}
\end{equation}
where \(\eta_K^2\) is a sufficiently large scaling factor that grows logarithmically with \(K\). \cite{Gutmann2016} \prettyref{fig:BOLFI} shows a visualization of the acquisition function, where \(\mu_K(\boldsymbol{\theta})\) is represented by the blue line and \(2\sqrt{\eta_K^2v_K(\boldsymbol{\theta})}\) is the width of the grey-shaded region around that blue line, while the orange line represents the acquisition function itself. \par
The minimum of the acquisition function approximates the parameter set with the highest chances of generating a simulation closest to the experiment. Hence, BOLFI generates samples sequentially and optimally to minimize the remaining uncertainty in the Likelihood estimation. \par
\begin{figure}[t]
    \includegraphics[width=\columnwidth]{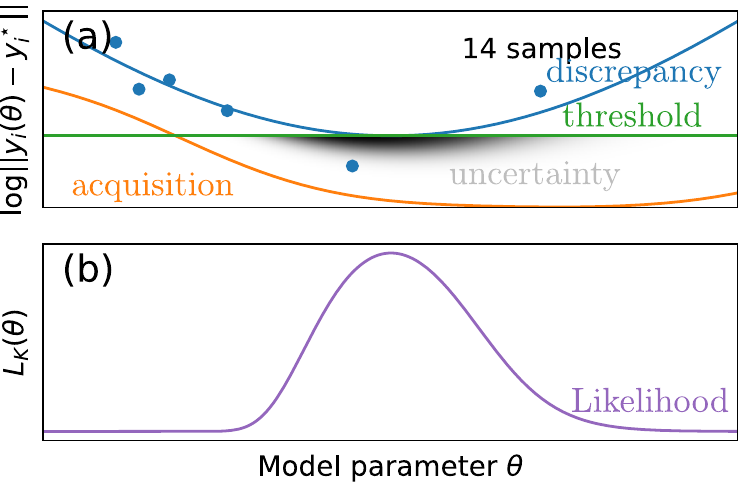}
    \caption{Calculation of the Likelihood approximation in BOLFI. \textcolor{defaultcolor}{\(L_K(\theta)\) is defined in \prettyref{eq:likelihood}.} (a) Cutout of \prettyref{fig:BOLFI} at 14 samples, zoomed in on the minimum of the discrepancy surrogate. The threshold \(\varepsilon\) is visualized as a green line. (b) The Likelihood is the integral of the discrepancy surrogate beneath the threshold.}
    \label{fig:Likelihood}
\end{figure}

After finishing the acquisition of samples, the final surrogate in \prettyref{eq:surrogate} forms the basis for the Likelihood approximation, as shown in \prettyref{fig:Likelihood}. In the Likelihood-Free Inference (LFI) framework, the approximation \(L_K\) to the Likelihood is the probability that the discrepancy surrogate falls below a threshold value \(\varepsilon\):
\begin{equation}
    L_K :\propto P(\log(\|y_i(\cdot)-y_i^\star\|) \leq \varepsilon). \label{eq:likelihood}
\end{equation}
The threshold value is arbitrary in most LFI algorithms, but BOLFI infers it from the surrogate itself as \(\varepsilon := \min_{\boldsymbol{\vartheta}}(\mu_K(\boldsymbol{\vartheta}))\). With the probability density function of the Gaussian surrogate, \(L_K\) can be calculated as
\begin{subequations}
\begin{align}
    L_K(y_i|\boldsymbol{\theta}) &\propto \int_{-\infty}^\varepsilon \exp\left( -\frac{(x - \mu_K(\boldsymbol{\theta})^2}{2(v_K(\boldsymbol{\theta}) + \sigma^2)} \right)\,dx\\
    &\propto F\left( \frac{\min_{\boldsymbol{\vartheta}}(\mu_K(\boldsymbol{\vartheta})) - \mu_K(\boldsymbol{\theta})}{\sqrt{v_K(\boldsymbol{\theta}) + \sigma^2}} \right),
\end{align}
\end{subequations}
where \(F\) is the cumulative distribution function of a standard normal variable,
\begin{equation}
    F(x) = \int_{-\infty}^x \frac{1}{\sqrt{2\pi}} \exp\left(-\frac{1}{2}u^2\right)\,du.
\end{equation}
%
%BOLFI convergence
Simultaneously regressing and sampling from the sampling distribution reduces the number of required samples by several orders of magnitude. \cite{Gutmann2016} \par

\section*{Validation of EP-BOLFI performance} \label{sec:performance}

\begin{table*}[t]
\begin{center}
\caption{Performance comparison between MCMC \cite{Aitio2020} and EP-BOLFI. The estimation results refer to one standard deviation.}
\label{tab:performance}
\begin{tabular}{cgcgc|cgcgcgc}
          & true  & initial std. & Aitio et al. (MCMC),         & EP-BOLFI & EP-BOLFI & EP-BOLFI \\
parameter & value & deviation        & \(\num{100000}\) simulations & \(\num{2080}\) sim. & \(\num{4160}\) sim. & \(\num{6240}\) sim. \\
\hline
\(D_e^*\)  /  \(\SI{e-10}{\square\metre\per\second}\) & \(\num{2.8}\) & \(\num{1.54}\) & \(2.8 \pm 0.007\phantom{0}\) & \(2.72 \pm 0.110\) & \(2.79 \pm 0.027\) & \(2.80 \pm 0.024\) \\
\(t_+\)  /  - & \(\num{0.4}\) & \(\num{0.156}\) & \(0.4 \pm 0.0008\) & \(0.42 \pm 0.012\) & \(0.40 \pm 0.003\) & \(0.40 \pm 0.003\) \\
\textcolor{defaultcolor}{\(D_{s,n}^*\)}  /  \(\SI{e-14}{\square\metre\per\second}\) & \(\num{3.9}\) & \(\num{1.39}\) & \(3.9 \pm 0.0005\) & \(3.91 \pm 0.158\) & \(3.90 \pm 0.021\) & \(3.90 \pm 0.004\) \\
\textcolor{defaultcolor}{\(D_{s,p}^*\)}  /  \(\SI{e-13}{\square\metre\per\second}\) & \(\num{1.0}\) & \(\num{1.98}\) & \(1.0 \pm 0.0002\) & \(1.09 \pm 0.167\) & \(1.00 \pm 0.019\) & \(1.00 \pm 0.005\) \\
\(\sigma^2\)  /  \(\SI{e-9}{\square\volt}\) & \(\num{1.6}\) & \(\num{1.17}\) & \(1.4 \pm 3.5\phantom{000}\) & \(2.75 \pm 0.40\phantom{0}\) & \(2.65 \pm 0.20\phantom{0}\) & \(1.66 \pm 0.09\phantom{0}\) \\
\end{tabular}
\end{center}
\end{table*}

%model and varying parameters
Our reference for state-of-the-art automated battery model parameterization is the work of Aitio et al. \cite{Aitio2020}. They create two types of synthetic data with an SPMe model, multimodal sinusoidal excitations at eleven SOCs denoted \textquote{excitation-point case} and a discharge with a superimposed small unimodal sinusoidal excitation denoted \textquote{wide-excursion case}. Aitio et al. fit five parameters with an MCMC algorithm: the electrolyte \(D_e^*\) and solid diffusivities \textcolor{defaultcolor}{\(D_{s,n}^*\), \(D_{s,p}^*\)}, the cation transference number \(t_+\), and the variance of the white noise superimposed on the synthetic measurement. In the wide-excursion case, MCMC fits the parameters nicely. However, the MCMC algorithm finds a wide range of inconsistent values in the excitation-point case \cite{Aitio2020}. \par
%output
We apply our EP-BOLFI algorithm to the same synthetic data with the same SPMe model as Aitio et al. \cite{Aitio2020}. Aitio et al. take the \(L^2\)-distance of the voltage as the cost function. For the EP-BOLFI algorithm, we preprocess the voltage response and define features. For the excitation-point cases, we perform a Fast Fourier Transform on the current input and voltage output to calculate the impedance for each mode as features. For the wide-excursion case, we use the \(L^2\)-distances of four voltage curve segments as features.  \par
%results
The results for the wide-excursion case are shown in \prettyref{tab:performance}. We observe that EP-BOLFI reaches a similar accuracy to that in Aitio et al. \cite{Aitio2020} with about 12 times less simulations, depending on the model parameter. Between \(\num{3120}\) and \(\num{6240}\) simulations for EP-BOLFI we observe an order of magnitude increase in accuracy, and at \(\num{8320}\) simulations all but \textcolor{defaultcolor}{\(D_{s,p}^*\)} are as accurately or even more accurateley estimated as by MCMC. \par
%direct comparison
The comparison of EP-BOLFI to MCMC in Aitio et al. \cite{Aitio2020} for the excitation-point case is shown in \prettyref{SI-tab:comparison}. Across all excitation points, EP-BOLFI with \(\num{6240}\) simulations vastly outperforms the MCMC approach in terms of stability and accuracy. \par
%further experiment
EP has the potential to deal with the data at all excitation points at once. Hence, we perform an additional experiment, collating the four excitation points 1, 2, 3 and 7 with the smallest uncertainties. While that leads to overfitting, the number of simulations required to reach convergence even shrinks to \(\num{2080}\). This demonstrates that the computation time of EP scales favourably with the dimension of the data \cite{Barthelme2015}. \par

\section*{Experimental} \label{sec:Experimental}

\subsection*{Experimental setup and a priori known parameters} \label{subsec:Setup}

\begin{table}[t]
\begin{center}
\caption{The missing battery parameters and how we procure them. For the symbols, cf. \prettyref{tab:parameters}.}
\label{tab:missing_parameters}
\begin{tabular}{c|cc}
parameter & source & value \\
\hline
\multicolumn{3}{l}{\textbf{A priori assumed parameters}} \\
\(\sigma_p^*\,[\si{\per\ohm\and\metre}]\) & value for NMC-111 \cite{Danner2016} & 1.07 \\
\(\sigma_n^*\,[\si{\per\ohm\and\metre}]\) & value for graphite \cite{Danner2016} & 10.67 \\
\(\alpha_{k,\ell}\) & symmetry assumption & 0.5 \\
\hline
\multicolumn{3}{l}{\textbf{\prettyref{subsec:OCV}}} \\
\(U_n^*\,[\si{\volt}]\) & GITT measurement \cite{Birkl2015} & [0.0, 1.0] \\
\(U_p^*\,[\si{\volt}]\) & our GITT measurement & [3.0, 5.0] \\
\(R_p^*\,[\si{\metre}]\) & scaling factor, cf. \prettyref{eq:lumped_parameters} \cite{Danner2016} & \(\num{5.5e-6}\) \\
\(R_n^*\,[\si{\metre}]\) & scaling factor, cf. \prettyref{eq:lumped_parameters} \cite{Danner2016} & \(\num{12e-6}\) \\
\(a_p^*\,[\si{\per\metre}]\) & scaling factor, cf. \prettyref{eq:lumped_parameters}  \cite{Mistry2021} & \(\frac{3 (1 - \varepsilon_p)}{R_p^*}\) \\
\(a_n^*\,[\si{\per\metre}]\) & scaling factor, cf. \prettyref{eq:lumped_parameters}  \cite{Mistry2021} & \(\frac{3 (1 - \varepsilon_n)}{R_n^*}\) \\
\hline
\multicolumn{3}{l}{\textbf{\prettyref{subsec:parameterization}}} \\
\(D_{s,p}^*\,[\si{\square\meter\per\second}]\) & EP-BOLFI fit & TBD \\
\(D_{s,n}^*\,[\si{\square\meter\per\second}]\) & EP-BOLFI fit & TBD \\
\(i_{se,p,0}^*\,[\si{\ampere\per\square\meter}]\) & EP-BOLFI fit & TBD \\
\(i_{se,n,0}^*\,[\si{\ampere\per\square\meter}]\) & EP-BOLFI fit & TBD \\
\(t_+\,[-]\) & EP-BOLFI fit & TBD \\
\(\beta_p\,[-]\) & EP-BOLFI fit & TBD \\
\(\beta_n\,[-]\) & EP-BOLFI fit & TBD
\end{tabular}
\end{center}
\end{table}

%origin and gotchas of the experimental data
The full cell GITT data got measured at BASF. Here, we list the parameters known before starting our estimation algorithms, following the checklist in \cite{Sun2021}. The only points in that checklist we do not fulfill are \textquote{specifications of used materials} and \textquote{coulombic efficiency}. \par
%experimental conditions
The GITT measurement protocol consists of repeating sets of \(\SI{360}{\s}\)-\(\SI{0.1}{\C}\), \(\SI{180}{\s}\)-\(\SI{0.2}{\C}\), \(\SI{72}{\s}\)-\(\SI{0.5}{\C}\) and \(\SI{36}{\s}\)-\(\SI{1.0}{\C}\) pulses with occasional \(\SI{30}{\s}\)-\(\SI{2.5}{\C}\) pulses each 25\% SOC (State-Of-Charge). \(\SI{1}{\C}\) corresponds to a theoretical capacity of \(\SI{0.03}{\ampere\hour}\). The rest periods at zero current between the pulses were \(\SI{15}{\min}\), with the exception of the \(\SI{2.5}{\C}\) pulses, which were enclosed in \(\SI{30}{\min}\) rests. The GITT experiment was conducted at 25 °C. The minimum and maximum operating voltage of the cell are \(\SI{2.7}{\volt}\) and \(\SI{4.2}{\volt}\), respectively. The GITT data spans from SOC 25\% to 100\%, i.e., 25\% discharged to 100\% discharged. \par
%battery materials
The cell contains the following materials. The negative electrode mostly consists of graphite with 95.7\% active material and is slightly overbalanced. The separator is Celgard 2500 with \(\SI{25}{\micro\m}\) thickness. The positive electrode is 94\% NCM-851005 with 3\% Solef 5130, 2\% SFG 6L and 1\% Super C 65. The electrolyte is \(\SI{1}{\mole\per\litre}\) LiPF\({}_6\) in EC:DEC 3:7 with 2\% VS as SEI former. \par
%known battery parameters
Here, we list the geometric parameters and the electrolyte properties. The cell is a square pouch cell. The cross-sectional areas are \(\SI{52}{\milli\m}\times\SI{52}{\milli\m}\), \(\SI{55}{\milli\m}\times\SI{55}{\milli\m}\), and \(\SI{50}{\milli\m}\times\SI{50}{\milli\m}\) for the negative electrode, separator, and positive electrode, respectively. The porosities are roughly 0.27, 0.55, and 0.29 for the negative electrode, separator, and positive electrode, respectively. The porosities of the electrodes were approximated by comparing their density with the bulk densities of graphite at \(\SI{2.26}{\g\per\centi\m\cubed}\) and NCM-851005 at \(\SI{4.8}{\g\per\centi\m\cubed}\) (times 94\%), respectively. The porosity of the separator and its Bruggeman coefficient of 3.6 were taken from Patel et al. \cite{Patel2003}. The thicknesses of the electrodes are \(\SI{45}{\micro\m}\) and \(\SI{25}{\micro\m}\) for the negative and positive electrodes, respectively. Both thicknesses were calculated from densities and areal mass loadings\textcolor{defaultcolor}{, i.e. the mass of active material per cross-sectional area of the respective current collector}. The areal mass loadings of the negative and positive electrode are \(\SI{0.072}{\kilogram\per\square\meter}\) and \(\SI{0.080}{\kilogram\per\square\meter}\), respectively. Thus, the electrolyte volume is about \(\SI{0.0926}{\milli\liter}\) and the volumetric ratio of electrolyte to active material is about 0.737. The electrolyte has cation transference number \(\num{0.3 +- 0.1}\), thermodynamical factor \(\num{1.475} / (1 - t_+)\), diffusivity \(\SI{3.69e-10}{\square\m\per\s}\), and conductivity \(\SI{0.950}{\per\ohm\per\m}\). We take the a priori electrolyte parameters from Nyman et al. \cite{Nyman2008}, even though they characterized EC:EMC 3:7, not EC:DEC 3:7. We neglect this disparity, since our focus is to demonstrate our parameterization procedure. We consider the cation transference number unknown with known error bounds and tie the thermodynamical factor to it as \(\num{1.475} / (1 - t_+)\). \par
%unimportant unknowns
At this point, we fixed most parameters. We discuss the remaining unknown parameters listed in \prettyref{tab:missing_parameters} in the next subsections. We shortly discuss the few unknown parameters with negligible impact here. The electronic conductivities of the electrodes are from Danner et al. \cite{Danner2016}; there, the electrode materials were graphite and NMC-111. We accept the potential error due to the different materials, since the electronic conductivities typically only account for a static IR drop of up to \SI{2}{\milli\volt} at current \SI{1}{\C}. The charge-transfer coefficients, also known as asymmetry factors, are set to \(0.5\), i.e., we make the standard assumption that the charge-transfer reactions are symmetric.  \par

\subsection*{Measurement of the OCV curves} \label{subsec:OCV}

%importance of OCV precision
The Open-Cell Voltage (OCV) has by far the most significant effect on the overall operating cell voltage, typically about \(\SI{3}{\V}\) to \(\SI{4.5}{\V}\). The minor voltage losses described by the equations in \prettyref{eq:nodim} are due to the limiting transport processes and reactions the so-called \textquote{overpotential}. The overpotential is typically around \(\SIrange{10}{100}{\milli\V}\). Hence, before we analyze cell processes via the overpotential, we need to know the OCV of the cell with high precision \cite{Huang2020}. \par
%OCV precision in y-axis: GITT
The most precise measurement of the OCV is possible with GITT, with a voltage error of \(\SIrange{0.1}{1}{\milli\V}\) \cite{Birkl2015}. Still, measurements at low constant current get often misused as \textquote{quasi-OCV} (qOCV) measurements \cite{Tang2019, Park2018}. Chen et al. \cite{Chen2020} demonstrate how constant-current curves exhibit significant deviations from OCV at high or low charge and between voltage plateaus, even at low current. Hence, we do not use qOCV measurements. \par
%OCV precision in x-axis: fitting function
We use the physics-inspired OCV model of Birkl et al. \cite{Birkl2015}, as it assigns an SOC range to incomplete OCV data in an informed and automatic way. This SOC range will approximate the maximally lithiated and de-lithiated states of the positive electrode at SOCs 0 and 1, respectively. We denote that SOC range as \textquote{positive electrode SOC} or \(\SOCp\). Likewise, we refer to the maximally lithiated and de-lithiated states of the negative electrode when we assign the \textquote{negative electrode SOC} or \(\SOCn\). \par
%theoretical capacity concentrations and OCV curves
We fit the OCV curve of the positive electrode given the OCV curve of the negative electrode in \prettyref{subsec:OCV_extraction}. The maximum intercalation concentrations were fitted to the CC-CV and GITT data at the same time. \par

\subsection*{Parameters fitted with EP-BOLFI} \label{subsec:parameterization}

\begin{figure}[t]
    \includegraphics[width=\columnwidth]{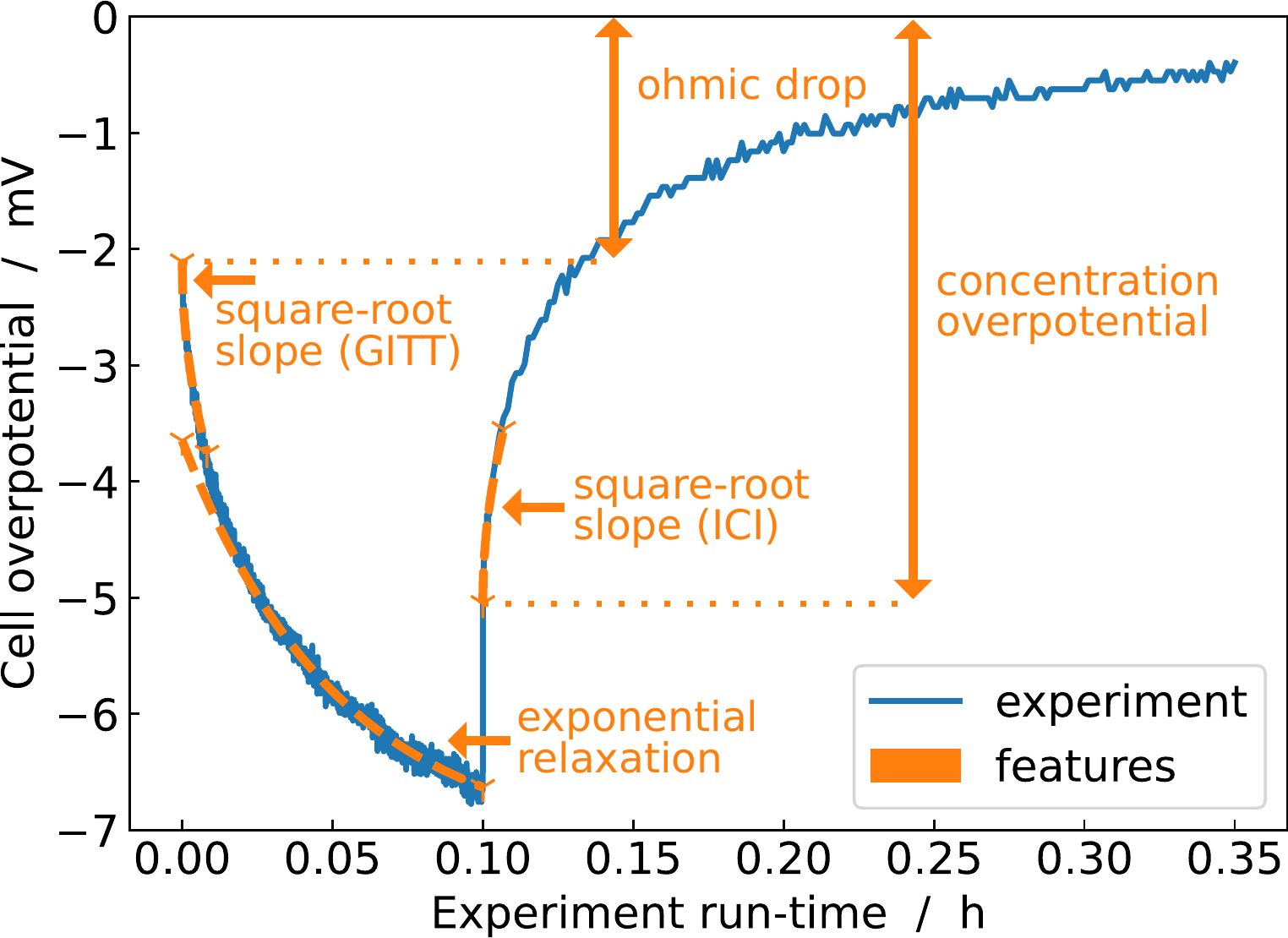}
    \caption{The five features \(y_i\) for Expectation Propagation in the experimental data at each GITT pulse.}
    \label{fig:gitt_features}
\end{figure}

%simulation task
We now have seven model parameters \(\boldsymbol{\theta}\) of interest that we can fit to the GITT data, summarized in \prettyref{eq:lumped_parameters}. These are the model parameters that appear independently from unknown properties in our physics-based model,
\begin{equation}
\boldsymbol{\theta}\subseteq\left\{\frac{D_{s,p}^*}{{R_p^*}^2},\ \frac{D_{s,n}^*}{{R_n^*}^2},\ i_{se,p,0}^* a_p^*,\ i_{se,n,0}^* a_n^*,\ t_+,\ \beta_n,\ \beta_p\right\}. \label{eq:lumped_parameters}
\end{equation}
%
%spatial homogenity
For simplicity, we assume that these parameters are independent from electrolyte concentrations and location within the cell. We ignore the concentration dependence of the cation transference number \cite{Landesfeind2019}. We assume that the Bruggeman coefficients are spatially constant. Spatially and concentration resolved measurements are required to parameterize these heterogeneities. \par
%SOC-dependency
GITT measurements give us information about SOC dependence. The functional form of the exchange-current density\\\textcolor{defaultcolor}{\(i_{se,k,0}^*(c_e^*, c_{s,k}^*)\)} is an active topic of research \cite{Colclasure2010, Latz2013, Kurchin2020, Fraggedakis2021}. Likewise, the active material diffusivities \textcolor{defaultcolor}{\(D_{s,k}^*(c_{s,k}^*)\)} depend on electrode SOC \cite{Xing2020}. Hence, there are three parameters of interest that do not depend on SOC, namely \(t_+\), \(\beta_p\), and \(\beta_n\), and four that do depend on SOC, namely \(i_{se,p,0}^*\), \(i_{se,n,0}^*\), \(D_{s,p}^*\) and \(D_{s,n}^*\). Therefore, we fit all seven parameters of interest in \prettyref{eq:lumped_parameters} once to a pair of GITT pulses with different C-rates. Then, we hold the three SOC-independent parameters constant, and fit only the four SOC-dependent ones to all other GITT pulses. \par
%lumped parameters
We also miss the geometric parameters for the electrode particles to correctly scale the active material diffusivities \(D_{s,k}^*\) and the exchange-current densities \(i_{se,k,0}^*\). However, their absolute values can be corrected when the specific surface areas and particle radii get measured. Without loss of generality, we assume the specific surface area \(a_k^* = 3 (1 - \varepsilon_k) / R_k^*\) \cite{Mistry2021}. Analogously, the particle radii are just a scaling factor in our model. We arbitrarily take the particle radii from a battery of Danner et al. \cite{Danner2016}. \par
%features
For Expectation Propagation, we reduce the experimental data to discrete features as visualized in \prettyref{fig:gitt_features}. The five features \(y_i\) we fit are the voltage directly after the current has been shut off (concentration overpotential), the relaxation times during the discharge pulses, the ohmic voltage drop, and the GITT and ICI (Intermittent Current Interruption) square-root slopes \cite{Chien2021} during discharge pulses and rest periods, respectively. \textcolor{defaultcolor}{We use the following fit function to obtain the square-root slopes \(\frac{dU}{d\sqrt{t}}\) and the offsets of the square-root segments \(U_0\):
\begin{equation}
    \text{square\_root\_fit}\left(t, U_0, \frac{dU}{d\sqrt{t}}\right) := U_0 + \frac{dU}{d\sqrt{t}} \cdot \sqrt{t - t_0},
\end{equation}
where \(t_0\) refers to the start of the current pulse or the start of the rest phase. The square-root function fitted to the current pulse gives the ohmic voltage drop as \(U_0\) and the GITT square-root slope as \(dU/d\sqrt{t}\). The square-root function fitted to the rest phase gives the concentration overpotential as \(U_0\) and the ICI square-root slope as \(dU/d\sqrt{t}\). To get the relaxation time \(\tau_r\), we use the following fit function on the current pulse:
\begin{equation}
    \text{exp\_fit}\left(t, U_0, \Delta U, \tau_r^{-1}\right) := U_0 + \Delta U \cdot \exp(-\tau_r^{-1} \cdot (t - t_0)), \label{eq:exponential_fit}
\end{equation}
where \(t_0\) refers to the start of the current pulse.}
% feature discussion
While our current pulses are too short to fulfill the requirements for ICI as an analytical formula, we will later see that our \textquote{incomplete} ICI square root slopes still give valuable information. We omit the exponential relaxation at rest as a feature since it is difficult to fit uniquely \cite{Bergstrom2021, Landesfeind2019}. This is discussed in \prettyref{sec:Discussions}. \textcolor{defaultcolor}{We omit \(U_0\) and \(\Delta U\) from the exponential fit function in \prettyref{eq:exponential_fit} as features, as they are less consistently fitted than the concentration overpotential, which contains much of the same information.} \par
% feature completeness
The analysis of GITT and ICI with approximate formulas introduces some inaccuracy as discussed by Geng et al. \cite{Geng2022}. EP-BOLFI fits the whole DFN model directly to the experimentally observed square root slopes. A detailed discussion for the complementary relevance of both GITT and ICI features is performed in \prettyref{SI-sec:features}. We emphasize that one might also use simple time segments of the data with their \(L^2\)-distance as features for EP-BOLFI if no better preprocessing step is known. The features we chose have less interdependence than \(L^2\)-segments, accelerating the computations and making the results more interpretable. \textcolor{defaultcolor}{As we can see in \prettyref{fig:gitt_features}, we capture the main information of the raw voltage curve in the five features.} \par
%probability distributions
We choose normal distributions for the Bruggeman coefficient priors and the cation transference number, and log-normal distributions for the priors of all other parameters of interest. We motivate the log-normal distributions with the Arrhenius relation: reaction rates and diffusivities may be modelled as a reaction following an Arrhenius relation, and hence are log-normally distributed if the corresponding activation energies are normally distributed. In the spirit of Laplace approximations \cite{Gelman2013}, a good approximation of the true distributions close to the true estimate is sufficient, even for global optimization. The bounds of the prior and posterior 95\% confidence intervals and the most likely estimates are listed in \prettyref{tab:bounds}. \par

\subsection*{Computational details} \label{subsec:setup}

%checklist
We employ the helpful checklist from Mistry et al. \cite{Mistry2021a} to ensure that we include every aspect of sensitivity of numerical inputs. The filled-in checklist is in \prettyref{SI-tab:checklist} in \prettyref{SI-sec:checklist}. \par
%discretization
We discretize the 1D+1D model equations in \prettyref{eq:nodim} with Spectral Volumes \cite{Wang2002} and solve them with CasADi \cite{Andersson2019} interfaced through PyBaMM \cite{Sulzer2020}. Compared to Finite Volumes and the ode15s solver in MATLAB, our simulations run 20 times faster. The Spectral Volumes mesh is of order 8 in the electrolyte and the negative electrode particles and of order 20 in the positive electrode particles. For the electrolyte mesh, 2 Spectral Volumes are in the negative electrode, and 1 Spectral Volume each is in the separator and the positive electrode. We check for mesh independence by running a simulation with orders 16 and 40 instead of 8 and 20, respectively: the features change by no more than 0.2\%. The timesteps are at most \(\SI{0.1}{\second}\), and reducing them to at most \(\SI{0.01}{\second}\) changes the features by no more than 2\%. \par
%fitting algorithms
Whenever we fit a parameterized function to data points, we use the trust-region algorithm for constrained optimization implemented in SciPy. The OCV model \cite{Birkl2015} we use gives the electrode SOC as a function of its OCV. Since the 1D+1D model equations in \prettyref{eq:nodim} require the inverse function, we invert OCV model fits numerically and fit a spline to the inverse. Compared to direct spline fits to data, we retain the model-informed smoothing of the data and the estimation of the SOC range. \par
%omitted uncertainties
Our algorithm can consider the uncertainties of all model parameters if the simulator function randomly samples them. Hence, we included the uncertainties of the SOC-independent fit parameters \(t_+\), \(\beta_n\) and \(\beta_p\) in the parameterization of the individual GITT pulses. \par
%iterative fitting procedure
For the initial fit of the SOC-dependent parameters, we choose the final GITT pulse in the dataset. The final GITT pulse corresponds to the most extreme SOC and, therefore, the largest overpotential, ideal for an initial fit. The SOC-dependent parameters \(i_{se,k}^*\) and \(D_{s,k}^*\) vary over two to four orders of magnitude, respectively. To allow for this parameter variability from one GITT pulse to the next, we magnify the error bars of the SOC-dependent parameters of one fit before we use them as Prior for the next fit. We limit the lower and upper bounds the priors could take to \([0.5, 80.0]\,\si{\ampere\per\square\meter}\) for the exchange-current densities and to \textcolor{defaultcolor}{[1e-14, 1e-10]}\(\,\si{\square\meter\per\second}\) for the electrode diffusivities. This limiter prevents the fit parameters from diverging into regions with infinite exchange-current density or instant diffusion. The error bars are then magnified to span at least half of the lower and upper bounds in the log scale. \par
%unknown model parameters
We fit the seven parameters of interest in \prettyref{eq:lumped_parameters} to GITT data in \prettyref{subsec:GITT} with our EP-BOLFI algorithm. We estimate \(n=4,7\) parameters with \(2^6+1, 2^7+1\) warm-up samples and \(2\cdot (2^6+1), 2\cdot (2^7+1)\) samples in total for BOLFI for each individual feature update, respectively. The pseudo-posteriors are estimated with an effective sample size of \(0.5\cdot6^2 + 1.5\cdot6,0.5\cdot7^2 + 1.5\cdot7\) for \(n=4,7\), respectively, by the No-U-Turn-Sampler \cite{Hoffman2014}, a Markov-Chain Monte-Carlo sampler implemented in ELFI \cite{Lintusaari2018}. We set up 4 EP iterations in both cases. We set the EP dampening parameter such that the total dampenings at the end are \(0.5\). \par
%verification
For validation of our novel EP-BOLFI algorithm, we repeat the estimation procedures with one difference. We replace the measured data with synthetic simulated data for the most likely fit parameters. The better a given parameter is identifiable from the features, the closer the verification run would be to the first results. With this validation, we also make sure that EP-BOLFI converged, such that we can be sure that invisible non-identifiability issues arise purely from model and data. \par
%computational time; one 4-parameters simulation took about 1.964 s, one 7-parameters simulation took about 4,3526 s
The total time for the estimation of seven parameters of interest with \(\num{10320}\) simulations on our office PC with an Intel Core i7-6700 at 4.00 GHz x 8 is about 20 hours. Of that, roughly twelve hours is the simulation time for \(\num{10320}\) simulations of two GITT pulses. The total time for the estimation of the four SOC-dependent parameters of interest with \(\num{2600}\) simulations for each pulse is about 120 minutes on our PC. Of that, roughly 80 minutes is the simulation time for \(\num{2600}\) simulations of one GITT pulse. \par

\section*{Results from full-cell GITT data} \label{sec:Results}

Before we describe the application of EP-BOLFI to GITT data, we have to identify the OCV of the positive electrode in \prettyref{subsec:OCV_extraction}. The application of EP-BOLFI is then presented in \prettyref{subsec:GITT}.

\subsection*{Extracting positive electrode OCV from full-cell GITT} \label{subsec:OCV_extraction}

\begin{figure}[t]
    \includegraphics[width=\columnwidth]{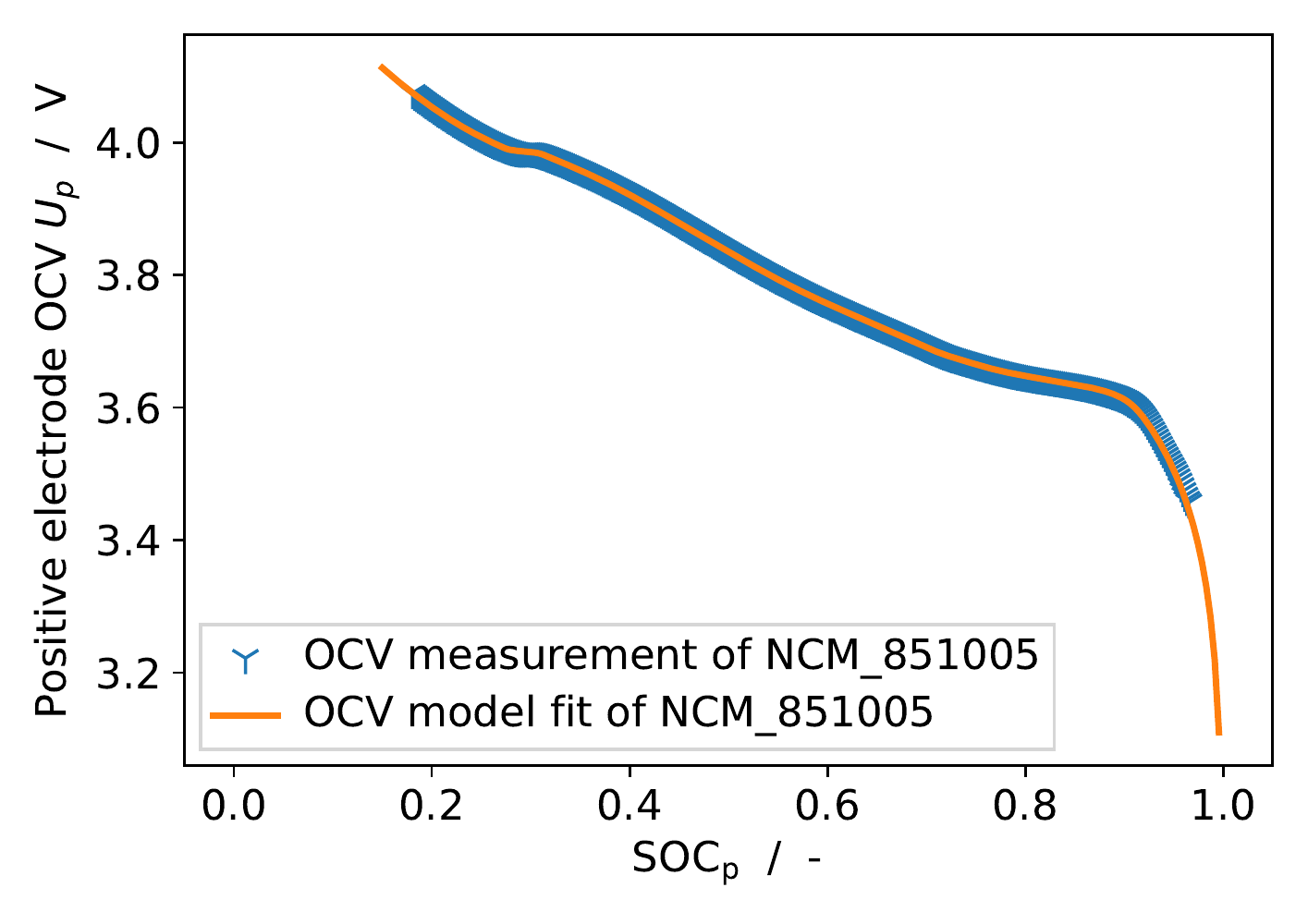}
    \caption{The model fit \cite{Birkl2015} of the OCV curve of the positive NCM-851005 electrode that we obtain by adding the OCV curve of the negative electrode to the GITT data. The root-mean-square-error is \(\SI{13}{\milli\volt}\) and the mean absolute error is \(\SI{0.25}{\milli\volt}\).}
    \label{fig:OCV_result}
\end{figure}

%negative electrode OCV alignment
We want to obtain the open-circuit voltage (OCV) \(U_p\) of the positive electrode. Since we do not have half-cell data, our estimation relies on our approximate knowledge of the OCV \(U_n\) of the negative electrode \cite{Birkl2015}.\par
%approximated second derivative from data
As a first step, we determine the cell balancing, i.e., the negative electrode SOC as a function of the positive electrode SOC. To find the cell balancing, we determine an approximate second derivative of \(U_n\) from the data. We obtain that by shifting the CC curves by the CV step against each other, adding them, and subtracting the GITT curve two times. The derivation of this preprocessing step and the utilized data are laid out in \prettyref{SI-sec:OCV_preprocessing}. \par
We compare the second derivative of \(U_n\) to that of a known graphite OCV curve \cite{Birkl2015} in \prettyref{SI-sec:OCV_preprocessing}. Though being a rough approximation, we can identify the peak positions of the second derivative. We find that the negative electrode SOC ranges from 3\% to 84\% over the SOC range of the positive electrode. \par
%positive electrode OCV
Taking into account this cell balance, we obtain \(U_p\) from the sum of \(U_n\) and the cell OCV from GITT data in \prettyref{fig:OCV_result}. We find that the cell capacity is \(\SI{39.65}{\milli\ampere\hour}\). The OCV model of Birkl et al. \cite{Birkl2015} with eight phases gets fitted to the SOC range \(0.18 \dots 0.97\) and we trust its extrapolation to the SOC range \(0.15 \dots 1.00\). We ignore OCV hysteresis effects \cite{Birkl2015} and only consider the discharge direction from now on. \par

\subsection*{Extracting model parameters with EP-BOLFI from GITT} \label{subsec:GITT}

\begin{table}[t]
\begin{center}
\caption{The prior 95\% bounds and posterior standard deviation bounds for the 7-parameter estimation, including the most likely estimate.}
\label{tab:bounds}
\begin{tabular}{c|cc}
Parameter & prior bounds & posterior estimate \\
\hline
\(i_{se,n,0}^*\,[\si{\ampere\per\square\metre}]\) & [1.0, 100.0] & 12.3, [10.2, 14.8] \\
\(i_{se,p,0}^*\,[\si{\ampere\per\square\metre}]\) & [1.0, 100.0] & 36.3, [29.9, 44.0] \\
\(D_{s,n}^*\,[\si{\square\metre\per\second}]\) & [1e-14, 1e-10] & (1.19, [0.82, 1.73])e-11 \\
\(D_{s,p}^*\,[\si{\square\metre\per\second}]\) & [1e-14, 1e-10] & (0.82, [0.56, 1.22])e-11 \\
\(t_+\,[-]\) & [0.2, 0.4] & 0.349, [0.343, 0.355] \\
\(\beta_n\,[-]\) & [1.8, 4.2] & 2.72, [2.64, 2.81] \\
\(\beta_p\,[-]\) & [1.8, 4.2] & 3.06, [2.99, 3.13]
\end{tabular}
\end{center}
\end{table}

\begin{figure}[t]
    \includegraphics[width=\columnwidth]{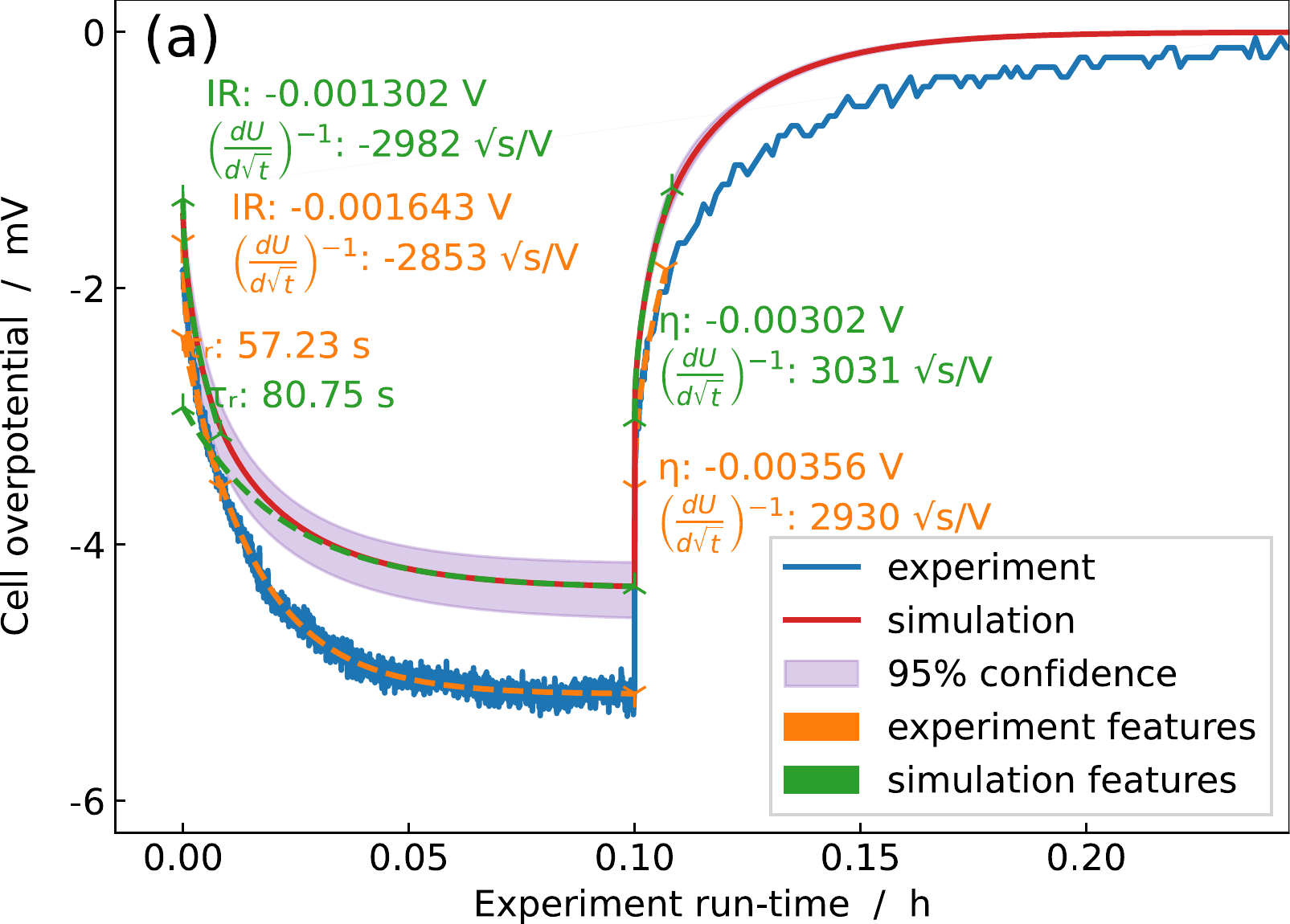}
    \includegraphics[width=\columnwidth]{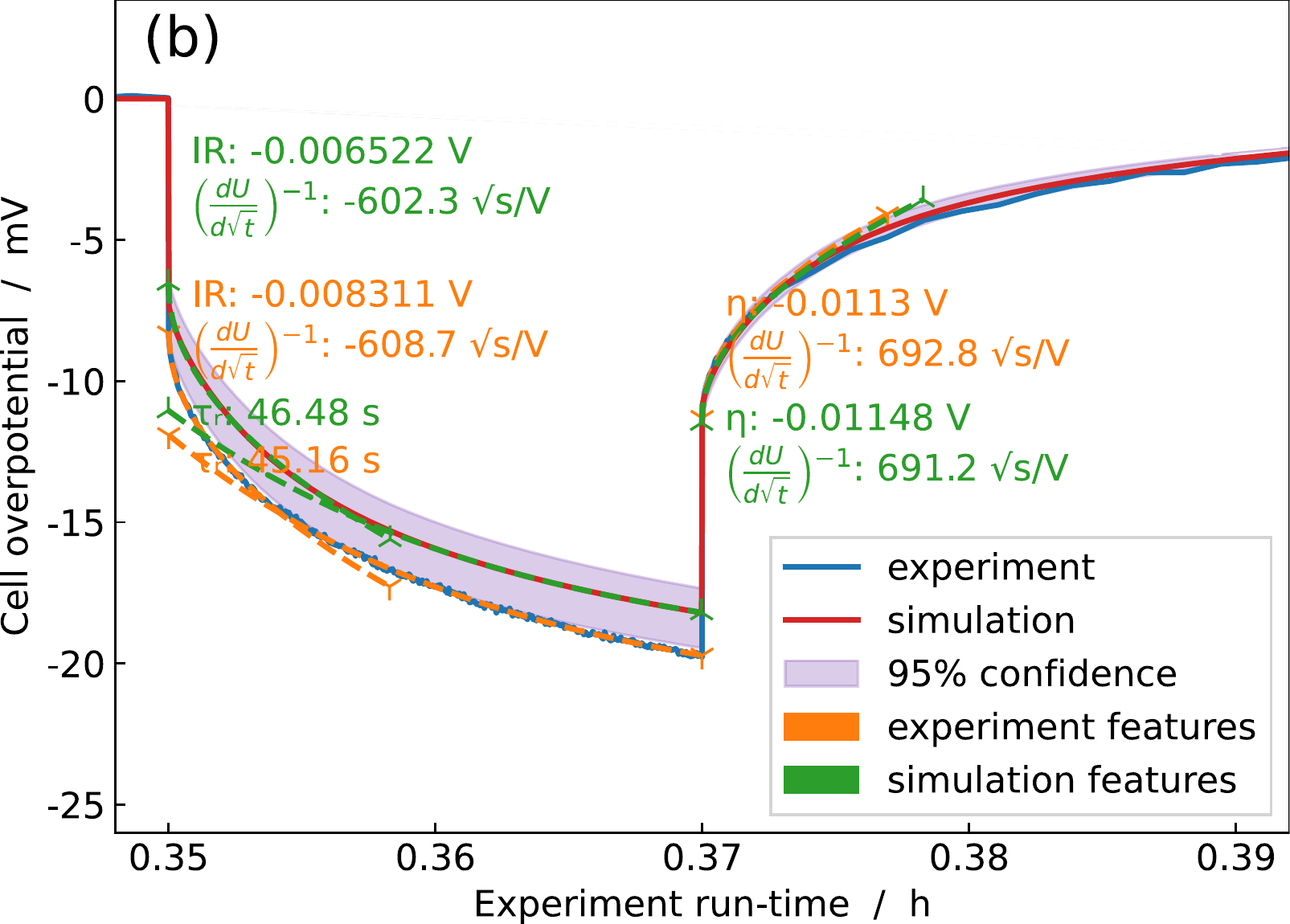}
    \caption{The two GITT pulses (a) 66 at \(\SI{0.1}{\C}\) and (b) 67 at \(\SI{1.0}{\C}\) to which we fitted all 7 model parameters of interest. We compare the features as presented in \prettyref{fig:gitt_features}.}
    \label{fig:gitt_measurement}
\end{figure}

\begin{figure}[t]
    \includegraphics[width=\columnwidth]{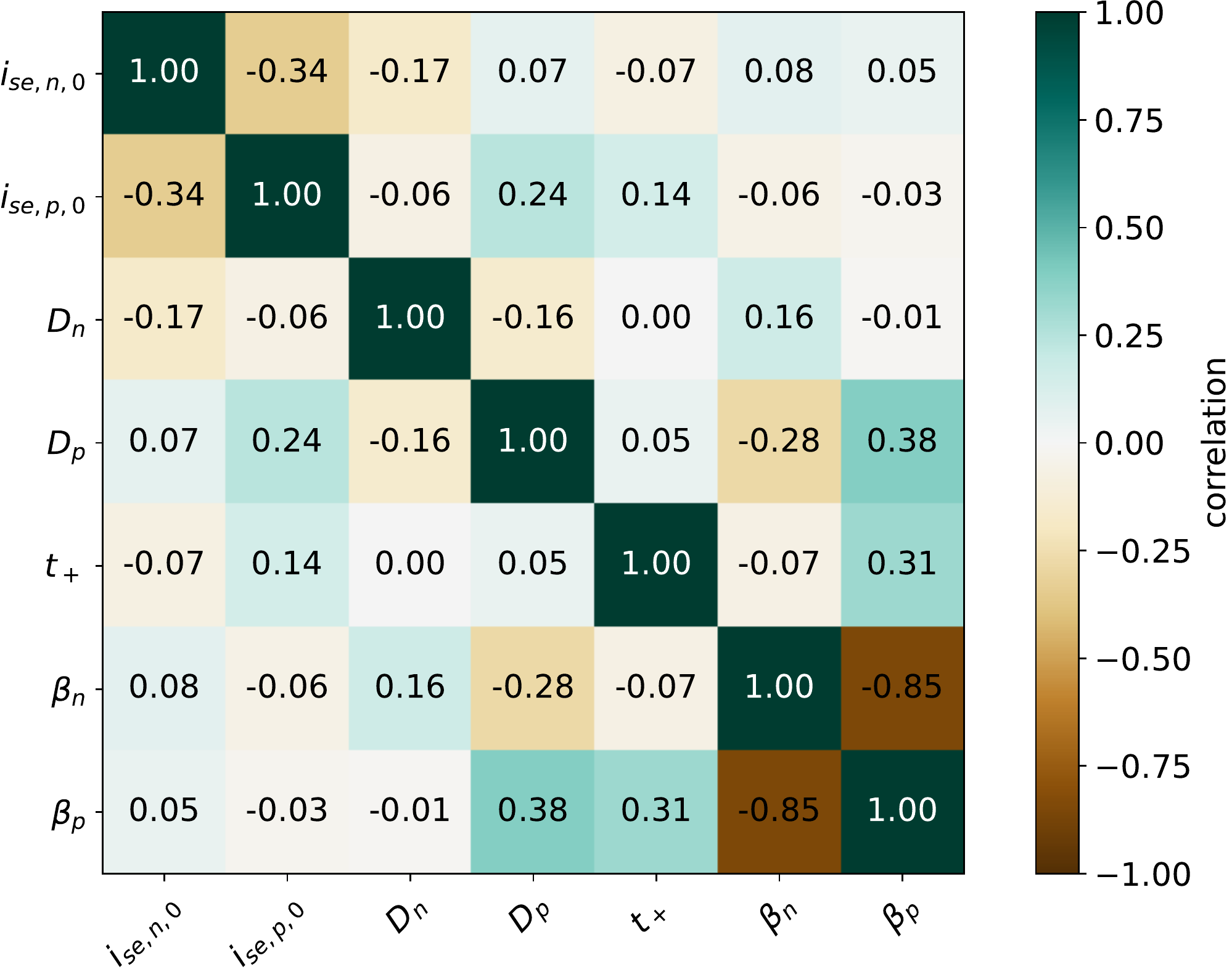}
    \caption{The correlation matrix of the estimation result for 7 model parameters from two GITT pulses.}
    \label{fig:correlation_gitt}
\end{figure}

\begin{figure*}[t]
    \includegraphics[width=\textwidth]{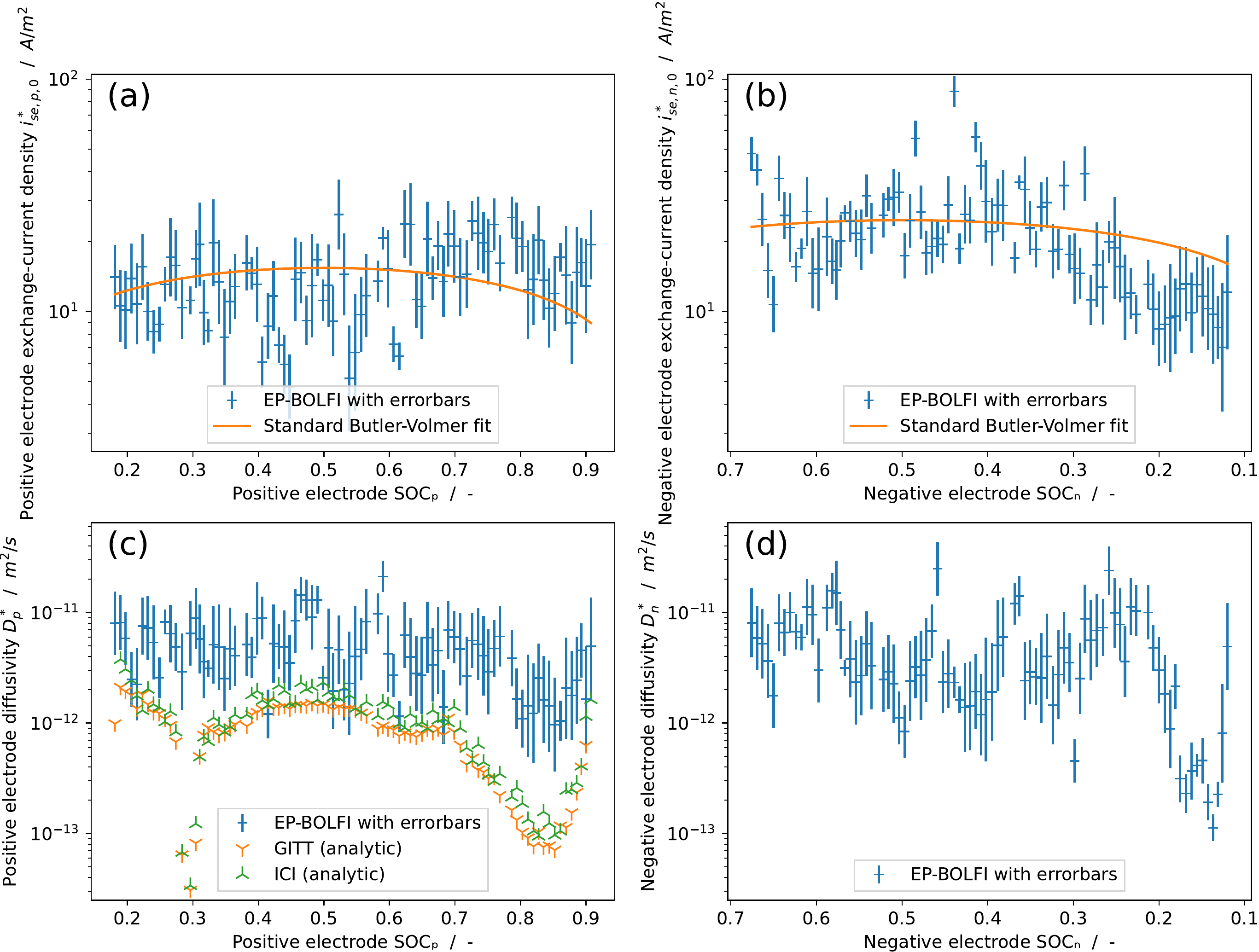}
    \caption{(a) \(i_{se,n,0}^*\), (b) \(i_{se,p,0}^*\), (c) \textcolor{defaultcolor}{\(D_{s,n}^*\)} and (d) \textcolor{defaultcolor}{\(D_{s,p}^*\)} estimates with error bars corresponding to one standard deviation of the logarithmic parameters. The comparison to a standard Butler-Volmer fit \(i_{se,k,0}^*(c_e^*, c_k^*=c_{s,k}^*|_{r_k^*=R_k^*}) = i_{se,k,0,0}^* \sqrt{c_e^*c_k^*} \sqrt{1 - c_k^*}\) (see \prettyref{eq:nodim-ise}) shows that we can not use the results for \(i_{se,n,0}^*\) and \(i_{se,p,0}^*\). However, if other parameters have been fitted correctly, their uncertainty then considers that \(i_{se,n,0}^*\) and \(i_{se,p,0}^*\) are effectively unknown. The results for \textcolor{defaultcolor}{\(D_{s,n}^*\)} are similar to those with half-cell GITT data from Schmalstieg et al. \cite{Schmalstieg2018}, Figure 7, apart from the wrong scaling. For \textcolor{defaultcolor}{\(D_{s,p}^*\)}, we plot the results of the analytic formulas for GITT and ICI for comparison. For \textcolor{defaultcolor}{\(D_{s,n}^*\) and \(D_{s,p}^*\)} we plot the corresponding OCV curves of their respective electrode.}
    \label{fig:ep_bolfi_fit}
\end{figure*}

\begin{figure*}
    \includegraphics[width=\textwidth]{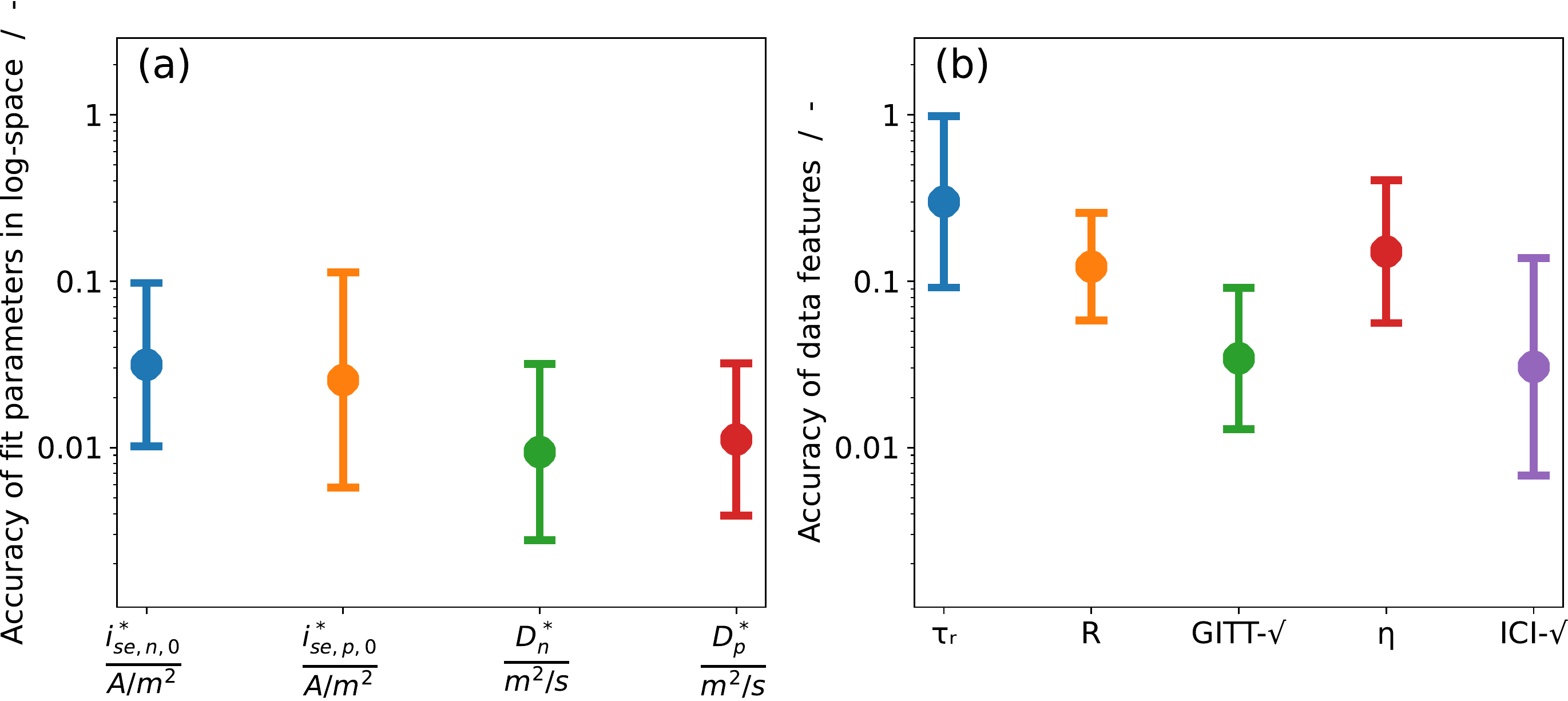}
    \caption{Summary plots where the circles indicate the mean and the bars indicate plus-minus one standard deviation. (a) The relative comparison of the verification run with the original fit parameters in log-space. We fit the DFN model to itself for each GITT pulse estimation to see if it identifies the same parameters. (b) The relative deviation of the fitted simulated features to the experimental features. We see that all features but the exponential relaxation time get close to the 2\% accuracy with which we simulate the features.}
    \label{fig:summary}
\end{figure*}

%seven-parameter estimation
With the procedure laid out in \prettyref{subsec:setup}, we fit seven parameters to GITT pulses 66 (\(\SI{0.1}{\C}\)) and 67 (\SI{1.0}{\C}) in the data. We choose these pulses since the OCV curves of both electrodes have significant gradients there, which improves the sensitivity of our experimental data to electrode diffusivities \cite{Chien2021}. Our prior bounds are based on the LiionDB database \cite{Wang2021} and an order-of-magnitude estimate of the exchange-current densities and active material diffusivities. We show the fitting results as estimates and 95\% confidence intervals in \prettyref{tab:bounds}. \par
%evaluation
In \prettyref{fig:gitt_measurement}, we depict the five fit features in the experiment and how we predict them in the simulation. The good agreement between features in simulation and data at low and high current verifies our optimization. The square-root GITT and ICI features have an excellent agreement. However, there is a significant difference in fit quality between low (\(\SI{0.1}{\C}\)) and high (\(\SI{1.0}{\C}\)) current for the exponential relaxation. We discuss that electrode heterogeneity can explain this mismatch (see \prettyref{sec:Discussions}). We find direct relationships between features and parameters below; thus, the excellent fit of most of the features gives insights on its own. \textcolor{defaultcolor}{\prettyref{fig:gitt_measurement}a is a good example to illustrate our featurization approach. If we had optimized the model for the time series, the simulation would cut across the voltage measurements, matching neither short-term nor long-term processes in the battery.} \par
%correlation
We now discuss the correlation matrix, i.e., the covariance matrix normalized by the variances. In \prettyref{fig:correlation_gitt}, the correlation matrix indicates how a parameter change affects the estimates of the other parameters. Each matrix entry shows how the (lack of) knowledge of some parameters influences the accuracy of the other parameters. Correlations close to 0 indicate that the two parameters are not coupled in the data as interpreted by our model. \textcolor{defaultcolor}{Note that} the two exchange-current densities are correlated with each other in our full cell examination. We see correlations between the SOC-independent model parameters and the SOC-dependent ones, especially between the diffusivity of the positive electrode and both Bruggeman coefficients. We rerun the estimation procedure twice with different initial random seeds to validate that these correlations are consistent properties of model and data rather than numerical artefacts. The corresponding data are in the accompanying GitHub repository. These correlations demonstrate the benefit of estimating all seven parameters together, namely that we obtain a consistent parameter set. \par
Next, we fit all GITT pulses with only the SOC-dependent four parameters \(i_{se,n,0}^*\), \(i_{se,p,0}^*\), \textcolor{defaultcolor}{\(D_{s,n}^*\) and \(D_{s,p}^*\)} under consideration of the remaining uncertainty of \(t_+\), \(\beta_n\) and \(\beta_p\). In \prettyref{fig:ep_bolfi_fit}, we plot the optimal estimates and error bars corresponding to one standard deviation of the logarithmic parameters as a function of electrode SOC. Please note that we define \textquote{electrode SOC} by the theoretical minimum and maximum lithiation of an electrode. The error bars in \prettyref{fig:ep_bolfi_fit} show us how precise the parameters are estimated in each GITT pulse.  \par
We compare the exchange-current densities in \prettyref{fig:ep_bolfi_fit}a and \prettyref{fig:ep_bolfi_fit}b with the standard SOC-dependence of the Butler-Volmer rate in \prettyref{eq:nodim-ise} \cite{Latz2013}. One reason for the large error bars is that the two exchange-current densities primarily act as interchangeable resistances. In \prettyref{SI-fig:joint_resistance}, we plot their joint resistance. Since it is not much smoother and has large error bars, we can conclude that there are more identifiability issues than just the interchangeability of the two exchange-current densities. \textcolor{defaultcolor}{The low identifiability is not surprising, as GITT was not designed with measuring exchange-current densities in mind. Both exchange-current densities are visible in measurements on much smaller timescales \cite{Heubner2016} than the GITT pulses used here, which means that they mostly collectively show up as a contribution to the ohmic drop.} \par
%\prettyref{SI-eq:gitt_ici_analytic}
The diffusivities in \prettyref{fig:ep_bolfi_fit}c and \prettyref{fig:ep_bolfi_fit}d are smoother functions of SOC. \textcolor{defaultcolor}{The error bars are still more prominent than in a half-cell setup, for which GITT was initially designed \cite{Weppner1977}. We attribute the larger error bars to a limited discernability of the two electrodes in the data.} To discuss the results for \textcolor{defaultcolor}{\(D_{s,p}^*\)}, we plot the results of the analytic formulas for the GITT and ICI methods \cite{Chien2021} for \textcolor{defaultcolor}{\(D_{s,p}^*\)} in \prettyref{fig:ep_bolfi_fit}c. We show these formulas in the SI Eq. 8.37. These formulas neglect the influence of the negative electrode and the electrolyte; hence, there is no equivalent formula for \textcolor{defaultcolor}{\(D_{s,n}^*\)}. We instead compare \textcolor{defaultcolor}{\(D_{s,n}^*\)} with a measurement by Schmalstieg et al. \cite{Schmalstieg2018}. \par
The results for \textcolor{defaultcolor}{\(D_{s,p}^*\)} in \prettyref{fig:ep_bolfi_fit}c have fairly small error bars and seem to be rather constant, in contrast to the analytical GITT and ICI formulas. The analytical formulas have especially large deviations from our results in the regions around \(\SOCp \approx 0.3\) and \(\SOCp \approx 0.85\). At \(\SOCp \approx 0.3\) the positive electrode OCV curve has a very flat plateau that impedes the analytic formula and introduces large uncertainty to it. At \(\SOCp \approx 0.85\) the low diffusivity of the negative electrode at corresponding \(\SOCn \approx 0.15\) may disturb the analytical formulas. EP-BOLFI, instead, produces consistent estimates for both diffusivities and can even distinguish between the two electrodes. Hence, the prediction of the uncertainty displayed in \prettyref{fig:ep_bolfi_fit}c gives us further information about the completeness of the experimental data and the predictability of the model parameters. \par
The results for \textcolor{defaultcolor}{\(D_{s,n}^*\)} in  \prettyref{fig:ep_bolfi_fit}d exhibit slightly smaller error bars than those for \textcolor{defaultcolor}{\(D_{s,p}^*\)} and reproduce the literature data of Schmalstieg et al. \cite{Schmalstieg2018} well. They observe the same shape of the diffusivity curve in a half-cell with graphite, especially the dip around \(\SOCn \approx 0.45\). Kang et al. \cite{Kang2021} perform a detailed analysis of the possible sources of uncertainty, such as the duration or intensity of the current pulses. \par
Please note that we can not determine the absolute magnitude of the estimates for \textcolor{defaultcolor}{\(D_{s,n}^*\) and \(D_{s,p}^*\)}, as they depend on the unknown particle radius \(R_n^*\) and \(R_p^*\), respectively. We emphasize that we can identify \textcolor{defaultcolor}{\(D_{s,n}^*\) and \(D_{s,p}^*\)} separately, as indicated by their weak correlation in \prettyref{fig:correlation_gitt} and their agreement with expected behaviours. We achieve this even though we measured a full cell and do not know the electrolyte properties perfectly. \par
In \prettyref{fig:summary}a we verify how reliable our parameterization result is. For this purpose, we create synthetic data from a DFN model with the parameters we fitted to the experimental data (see \prettyref{fig:gitt_measurement}). We apply EP-BOLFI to this SOC-dependent synthetic data in a verification run. We depict here the relative deviation between the true and estimated parameters in this verification run. As the fitted probability distributions are log-normal, we observe the relative deviations on a log-scale. We confirm that the solid diffusivities are identifiable within 1\% accuracy. Notably, the exchange-current densities show a much larger variability in accuracy, matching their erratic behaviour in \ref{fig:ep_bolfi_fit}ab. We conclude that the diffusivities in \prettyref{fig:ep_bolfi_fit}cd can be trusted at most SOC, while the exchange currents in \prettyref{fig:ep_bolfi_fit}ab are not clearly identifiable from GITT. \textcolor{defaultcolor}{We also verify that the model justifies the limitation to a Gaussian Posterior with only one mode since multiple distant sets of equally optimal parameters would result in far poorer accuracy.} \par
In \prettyref{fig:summary}b, we verify that the parameterization accurately reproduced the experimental data. For that, we collect the deviations between features in the experimental data to the features in the fitted simulated data. We see that all features were fitted well except for the exponential relaxation time \(\tau_r\). We conclude that \(\tau_r\) is an unreliable feature in our GITT dataset with the DFN model. \par
In \prettyref{SI-sec:sensitivity}, we perform a complementary sensitivity analysis on the DFN model at the fitted parameters at each GITT pulse. Based on these sensitivities, we can identify parameters that appear to be fitted well in \prettyref{fig:summary}a, but depend on unreliable features in \prettyref{fig:summary}b. As we can see in \prettyref{SI-fig:sensitivities}, \textcolor{defaultcolor}{\(D_{s,n}^*\) and \(D_{s,p}^*\)} both depend on all features but the ohmic resistance, with \textcolor{defaultcolor}{\(D_{s,n}^*\)} being more sensitive to the concentration overpotential than \textcolor{defaultcolor}{\(D_{s,p}^*\)}. We find that all four SOC-dependent parameters also have a consistently high sensitivity to the exponential relaxation time \(\tau_r\), which we just ruled out as a reliable feature. Despite that sensitivity, EP-BOLFI automatically ignored \(\tau_r\) in favour of fitting the parameters more consistently with other features. We conclude that we can nicely estimate \textcolor{defaultcolor}{\(D_{s,n}^*\) and \(D_{s,p}^*\)} accurately from our GITT dataset because they are sensitive to the square-root features and the concentration overpotential. \par

\section*{Discussions} \label{sec:Discussions}

%overview over the benefits of our approach/algorithm
We gain three benefits from our Bayesian method, EP-BOLFI. The first benefit is universality; treating the model as a black box allows us to change the equations and fit parameters arbitrarily. The second benefit is global optimization, i.e., the thorough exploration of the range of feasible parameters. The third benefit is that we consistently estimate both the fit parameters and their uncertainty with only one algorithm. \par
%state-of-the-art is gradient-based => not generalizable
State-of-the-art battery parameterization most often requires a complicated calculation of the parameter-output-gradient \cite{Sethurajan2019, Zhao2020}. For example, Sethurajan et al. \cite{Sethurajan2019} and Zhao et al. \cite{Zhao2020} conduct microscopic imaging of electrolyte concentration profiles. They then apply gradient-based iterative optimization to diffusion equations to extract electrolyte properties. While this specialization produces fantastic results, adapting their algorithm to other battery parameters or differential equations would warrant a new paper. With our algorithm, you only need to update the simulator and the feature definitions. \par
%efficient global optimization
The \textquote{exploration} aspect of BOLFI ensures that it evenly minimizes the remaining uncertainty across the whole parameter search space. A gradient-based approach would be limited to the estimate \textquote{closest} to the initial guess. Repeating the gradient-based optimization from various starting points can alleviate this locality problem, but it quickly becomes infeasible with a growing number of estimated parameters. \par
%why needing only one algorithm is not just easier, but also more powerful
The Bayesian analysis of Sethurajan et al. \cite{Sethurajan2019} and Zhao et al. \cite{Zhao2020} is preceded by a gradient-based optimization. EP-BOLFI is stable enough not to require this step, drastically cutting down on the implementation effort while incorporating a more expansive parameter space into the uncertainty estimate. \par
%optimization algorithm comparison to state-of-the-art
The comparable computation time for the simulations and the EP-BOLFI algorithm shows a considerable cost to optimizing the samples. Compared to Aitio et al. \cite{Aitio2020}, we require at least 12 times fewer samples for the estimation of five parameters. Hence, we significantly reduce the computation time and enable the simultaneous estimation of more than four parameters. \par
%goal: parameterization <-> characterization
The primary use of our technique is model parameterization, i.e., it tunes the model for predictive simulations, e.g., for a digital twin. Depending on the sophistication of the model, its parameters might be very close to the actual material properties. Thus, we may also characterize the battery non-destructively multiple times over its lifetime. Established characterization techniques usually require separate setups with Li-metal electrodes or destructive disassembly \cite{Weppner1977, Sethurajan2015, Chen2020}. Our results for \textcolor{defaultcolor}{\(D_{s,n}^*\)} are comparable to those of a half-cell GITT measurement \cite{Schmalstieg2018}, indicating that we extract the actual electrode property rather than some effective cell parameter. \par
%why do analytic formulas not suffice
We believe that directly fitting continuum models includes significant coupled effects that analytic approximations can not describe. For example, the analytical formulas for GITT date back to 1977 \cite{Weppner1977}, and are only accurate for electrodes whose Open-Circuit Voltage (OCV) is locally describable by a Nernst equation. Modern electrode materials feature OCV curves with plateaus and kinks that severely impact GITT measurements, as we saw in \prettyref{subsec:GITT} in \prettyref{fig:ep_bolfi_fit}c. Our approach is less \textcolor{defaultcolor}{sensitive to} this problem, as seen in \prettyref{fig:ep_bolfi_fit}c. \par
%how our algorithm even deals better with non-deliberately produced data; easiest use-case
Our focus is salvaging CC-CV and GITT measurements and other organically grown measurement protocols such as WLTP since they constitute most of the experimental data available. Hence, our algorithm can also deal with raw data that is simply divided into multiple time segments, where their respective \(L^2\)-distances between measurement and data would be the features. Such a segmentation reproduces the total \(L^2\)-distance while utilizing the benefits of Expectation Propagation. The result is a better global \(L^2\)-fit with less computational cost. This flexibility alleviates the dependence on data that is generated specifically for optimal parameter estimation performance \cite{Lai2021}. \par
%how our algorithm may help in designing an optimal estimation procedure
We can optimize a fitting procedure itself using the results of our algorithm. By studying the itemized correlation matrices for each feature, we can select the feature set with the clearest distinguishability between fit parameters. An analysis of the individual identifiability of each fit parameter is possible with either the diagonal of the covariance matrix or an interval analysis with the corresponding 95\% confidence intervals.  \par
%slow relaxation not reproduced well
The poor agreement between model and data exponential relaxation times at rest is likely due to the artificial homogeneity of the DFN model. The spatial heterogeneities of a battery cell significantly impact its performance and are best captured by microstructure-resolved modelling \cite{Less2012, Bolay2021}. The link between exponential relaxation times at rest and heterogeneity has been discussed by Kirk et al. \cite{Kirk2021}. They propose at least a multi-particle 1D+1D model, the MP-DFN, that incorporates the different length scales of the electrode particles. \par

\section*{Conclusion} \label{sec:Conclusion}

%you can use our algorithm to great avail
Our newly developed algorithm EP-BOLFI is an optimizer that only requires the to-be-optimized model while minimizing the algorithm setup and the model evaluations. At the same time, the results of EP-BOLFI grant further insight into the data and the measurement uncertainty. EP-BOLFI and the data we applied it to are freely available. The GitHub repository containing the algorithm and the data is linked in \prettyref{sec:Code}. \par
%recap of the paper
We show that EP-BOLFI is more robust than MCMC, while simultaneously being an order of magnitude faster. \textcolor{defaultcolor}{The segmentation into expert-informed features allows the algorithm to reliably match the model to the parts of the data that it can actually reproduce.} We successfully parameterize a DFN model to a real full-cell GITT experiment while treating the model as a black box. \textcolor{defaultcolor}{\textquote{Black box} refers to the fact that EP-BOLFI does not require a pre-calculated gradient of the model-data discrepancy with the model parameters.} Our results extract both electrode diffusivities, which would not be possible with the analytical GITT formula. \par
%what phenomenons our algorithm might be useful for
The non-destructive parameterization of models for the SEI \cite{Single2019}, double-layers \cite{Luck2019}, and plating \cite{Hein2020} might be possible with our algorithm. We expect to elucidate important correlations between these effects. \par
%benefit of our algorithm: combination of various measurements
EP-BOLFI allows to combine different measurement techniques for parameterisation. For example, Electrochemical Impedance Spectroscopy (EIS) can distinguish processes occurring at different time scales like reaction kinetics. This is a deficiency of GITT, as shown in this article. EP-BOLFI could fit EIS and GITT measurements simultaneously and determine a greater parameter set. \par
%use ROMC for something that more closely resembles Aitio et al. in terms of raw stability (many samples), while still using as few samples as possible
%slight suggestion to parallelize EP if raw computation power is more readily available than time
An adaptation of our algorithm to faster simulators or more parallel processing power may swap out BOLFI for Robust Optimization Monte Carlo (ROMC) \cite{Ikonomov2020}, implemented in ELFI \cite{Lintusaari2018}. For fitting a large number of features, EP may be (partially) parallelized. The EP features can be updated iteratively or in parallel. But they can alternatively be updated in a in-between manner, where one can still reap some of the benefits of iterative preconditioning while utilizing more parallel processing power \cite{Barthelme2015}. \par

\section*{Code availability} \label{sec:Code}

\urlstyle{same}
The bytecode of the presented EP-BOLFI optimization algorithm and the code which we use to create the figures is available at the following GitHub repository:\\\url{https://github.com/YannickNoelStephanKuhn/EP-BOLFI}.\\The source code of EP-BOLFI will be available at that same GitHub repository at a later date. The experimental data is there as well in the \textquote{Releases} section.

\begin{table*}[t]
\caption{Physical battery parameters and their non-dimensionalizations. \textcolor{defaultcolor}{\(j=\) s: active material phase. \(j=\) e: electrolyte. \(j=\) se: electrolyte-electrode interface. \(k=\) n/p: negative/positive electrode. \(k=\) s: separator. \(^*\): dimensional symbol. \(typ\) or \(ref\): arbitrary reference parameter.}}
\label{tab:parameters}
\begin{tabular}{cccc}
    Parameters & Units & Description & Non-dimensionalization \\
    \hline
    \multicolumn{4}{l}{\textbf{Spatial and temporal discretization}} \\
    \(L^*\) & \(\si{metre}\) & Thickness of the cell & \(L := 1\) \\
    \(L_k^*\) & \(\si{metre}\) & Thickness of cell component & \(L_k := L_k^* / L^*\) \\
    \(x^*\) & \(\si{metre}\) & Spatial coordinate through the cell & \(x := x^* / L^*\) \\
    \(R_k^*\) & \(\si{metre}\) & Radius of electrode particles & \(R_k := 1\) \\
    \(r_k^*\) & \(\si{metre}\) & Radial coordinate through a particle & \(r_k := r_k^* / R_k^*\) \\
    \(\tau_d^*\) & \(\si{\second}\) & Discharge timescale & \(\tau_d^* := F^* c_{s,p,max}^* L^* / I_{typ}^*\) \\
    \(t^*\) & \(\si{\second}\) & Time since beginning of experiment & \(t := t^* / \tau_d^*\) \\
    \hline
    \multicolumn{4}{l}{\textbf{Electrolyte variables}} \\
    \(c_e^*(t^*, x^*)\) & \(\si{\mole\per\cubic\metre}\) & Electrolyte concentration & \(c_e := c_e^* / c_{e,typ}^*\) \\
    \(\phi_{e,k}^*(t^*, x^*)\) & \(\si{\volt}\) & Electrolyte potential & \(\phi_{e,k} := \frac{\phi_{e,k}^* - U_{n,ref}^*}{R^* T^* / F^*}\) \\
    \(i_{e,k}^*(t^*, x^*)\) & \(\si{\ampere\per\square\metre}\) & Electrolyte ionic current density & \(i_{e,k} := i_{e,k}^* / I_{typ}^*\) \\
    \(N_{e,k}^*(t^*, x^*)\) & \(\si{\mole\per\square\metre\per\second}\) & Molar ionic \textcolor{defaultcolor}{flux} & \(N_{e,k} := \frac{N_{e,k}^* L^*}{D_{e,typ}^* c_{e,typ}^*}\) \\
    \multicolumn{4}{l}{\textbf{Electrode variables}} \\
    \(c_{s,k}(t^*, x^*, r^*)\) & \(\si{\mole\per\cubic\metre}\) & Concentration of intercalated lithium & \(c_{s,k} := c_{s,k}^* / c_{s,k,max}^*\) \\
    \(\phi_{s,k}^*(t^*, x^*)\) & \(\si{\volt}\) & Electrode potential & \(\phi_{s,k} := \frac{\phi_{s,k}^* - (U_{k,ref}^* - U_{n,ref}^*)}{R^* T^* / F^*}\) \\
    \multicolumn{4}{l}{\textbf{Electrolyte-Electrode interface reaction variables}} \\
    \(i_{s,k}^*(t^*, x^*)\) & \(\si{\ampere\per\square\metre}\) & Electrode electronic current density & \(i_{s,k} := i_{s,k}^* / I_{typ}^*\) \\
    \(\eta_k^*(t^*, x^*)\) & \(\si{\volt}\) & Reaction overpotential & \(\eta_k := \eta_k^* / (R^* T^* / F^*)\) \\
    \(i_{se,k}^*(t^*, x^*)\) & \(\si{\ampere\per\square\metre}\) & Intercalation reaction current density & \(i_{se,k} := a_k^* L^* i_{se,k}^* / I_{typ}^*\) \\
    \hline
    \multicolumn{4}{l}{\textbf{Operating conditions}} \\
    \(I^*(t^*)\) & \(\si{\ampere\per\square\metre}\) & Current density applied to the battery & \(I := I^* / I_{typ}^*\) \\
    \(T^*\) & \(\si{\kelvin}\) & Temperature of the cell & \(T := T^* / T_{ref}^*\) \\
    \multicolumn{4}{l}{\textbf{Electrolyte parameters}} \\
    \(D_e^*(c_e^*)\) & \(\si{\square\metre\per\second}\) & Electrolyte diffusivity & \(D_e(c_e) := D_e^* / D_{e,typ}^*\) \\
    \(\kappa_e^*(c_e^*)\) & \(\si{\per\ohm\per\metre}\) & Electrolyte ionic conductivity & \(\kappa_e := \kappa_e^*(c_{e,typ}^* c_e) / \kappa_{e,typ}^*\) \\
    \(t_+(c_e^*)\) & -- & Cation transference number & -- \\
    \(1 + \frac{\partial\ln(f_+)}{\partial\ln(c_e^*)}\) & -- & Thermodynamic factor & -- \\
    \multicolumn{4}{l}{\textbf{Electrode parameters}} \\
    \(\varepsilon_k\) & -- & Electrode porosity & -- \\
    \(\beta_k\) & -- & Electrode Bruggeman coefficient & -- \\
    \(D_{s,k}^*(c_{s,k}^*)\) & \(\si{\square\metre\per\second}\) & Electrode active material diffusivity & \(D_{s,k}(c_{s,k}) := D_{s,k}^* / D_{s,k,typ}^*\) \\
    \(\sigma_k^*\) & \(\si{\per\ohm\per\metre}\) & Electrode electronic conductivity & \(\sigma_k :=  \frac{R^*T^*}{F^*}\left/\left(\frac{I_{typ}^*L^*}{\sigma_k^*}\right)\right.\) \\
    \(a_k^*\) & \(\si{\per\metre}\) & Electrode surface area to volume ratio & \(a_k := R_k^* a_k^*\) \\
    \(U_k^*(c_{s,k}^*)\) & \(\si{\volt}\) & Open-Cell Voltage (OCV) & \(U_k := \frac{U_k^* - U_{k,ref}^*}{R^* T^* / F^*}\) \\
    \multicolumn{4}{l}{\textbf{Electrolyte-Electrode interface reaction variables}} \\
    \(\alpha_{(k=n,p),(k=n,p)}\) & -- & Reaction symmetry factors & -- \\
    \(z_k\) & -- & Charge transfer numbers & -- \\
    \(i_{se,k,0}^*(c_e^*, c_{s,k}^*)\) & \(\si{\ampere\per\square\metre}\) & Exchange-current density & \(i_{se,k,0} := i_{se,k,0}^* / i_{se,k,0,ref}^*\)
\end{tabular}
\end{table*}

\begin{table*}[t]
\caption{Scalings used for non-dimensionalization.}
\label{tab:scalings}
\begin{tabular}{cccc}
    Symbol & Units & Description & Definition \\
    \hline
    \(\tau_d^*\) & \(\si{\second}\) & Discharge timescale & \(F^* c_{s,p,max}^* L^* / I_{typ}^*\) \\
    \(\tau_e^*\) & \(\si{\second}\) & Electrolyte transport timescale & \({L^*}^2 / D_{e,typ}^*\) \\
    \(C_e\) & -- & Ratio of electrolyte transport and discharge timescales & \(\tau_e^* / \tau_d^*\) \\
    \(\gamma_e\) & -- & Ratio of electrolyte and electrode concentration & \(c_{e,typ}^* / c_{s,p,max}^*\) \\
    \(\hat{\kappa}_e\) & -- & Ratio of thermal voltage to ionic resistance & \(R^* T^* \kappa_{e,typ}^* / (F^* I_{typ}^* L^*)\) \\
    \(\tau_k^*\) & \(\si{\second}\) & Particle transport timescale & \({R_k^*}^2 / D_{s,k,typ}^*\) \\
    \(C_k\) & -- & Ratio of partical transport and discharge timescales & \(\tau_k^* / \tau_d^*\) \\
    \(\gamma_k\) & -- & Maximum electrode through maximum positive electrode concentration & \(c_{s,k,max}^* / c_{s,p,max}^*\) \\
    \(\tau_{r,k}^*\) & \(\si{\second}\) & Intercalation reaction timescale & \(F^* c_{s,k,max}^* / (i_{se,k,0,ref}^* a_k^*)\) \\
    \(C_{r,k}\) & -- & Ratio of reaction to discharge timescale & \(\tau_{r,k}^* / \tau_d^*\) \\
\end{tabular}
\end{table*}

\clearpage

\section*{Acknowledgements} \label{sec:Acknowledgements}

This work was supported by the German Aerospace Center (DLR). The authors thank BASF for providing the experimental data. The authors acknowledge support by the Helmholtz Association through grant no KW-BASF-6 (Initiative and Networking Fund as part of the funding measure \textquote{ZeDaBase-Batteriezelldatenbank}). This work contributes to the research performed at CELEST (Center for Electrochemical Energy Storage Ulm-Karlsruhe). \par

\section*{Conflict of Interest}

The authors declare no conflict of interest.

\section*{ORCID} \label{sec:ORCID}

\urlstyle{same}
\noindent
Yannick Kuhn \includegraphics[height=\baselineskip]{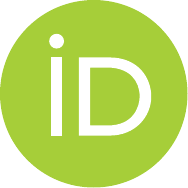} \url{https://orcid.org/0000-0002-9019-2290} \\
Birger Horstmann \includegraphics[height=\baselineskip]{Orcid_logo.pdf} \url{https://orcid.org/0000-0002-1500-0578} \\
Arnulf Latz \includegraphics[height=\baselineskip]{Orcid_logo.pdf} \url{https://orcid.org/0000-0003-1449-8172 } \\

%%%%%%%%%%%%%%%%%%%%%%%%%%%%%%%%%%%%%%%%%%%%%%%%%%%%%%%%%%
%%%%%%%%%%%%%%%%%%%%%%%%%%%%%%%%%%%%%%%%%%%%%%%%%%%%%%%%%%
%%%%%%%%%%%%%%%%%%%%%%%%%%%%%%%%%%%%%%%%%%%%%%%%%%%%%%%%%%
\begin{shaded}
\noindent\textsf{\textbf{Keywords:} \keywords} 
\end{shaded}
%%%%%%%%%%%%%%%%%%%%%%%%%%%%%%%%%%%%%%%%%%%%%%%%%%%%%%%%%%
%%%%%%%%%%%%%%%%%%%%%%%%%%%%%%%%%%%%%%%%%%%%%%%%%%%%%%%%%%
%%%%%%%%%%%%%%%%%%%%%%%%%%%%%%%%%%%%%%%%%%%%%%%%%%%%%%%%%%

%%%%%%%		References			%%%%%%% 

\setlength{\bibsep}{0.0cm}
\bibliographystyle{Wiley-chemistry}
\bibliography{main}

\begin{thebibliography}{10}

\bibitem{Latz2011}
A.~Latz, J.~Zausch, \emph{J. Power Sources} \textbf{2011}, \emph{196}, 3296.

\bibitem{Fong2021}
K.~D. Fong, J.~Self, B.~D. McCloskey, K.~A. Persson, \emph{Macromol.}
  \textbf{2021}, \emph{54}, 2575.

\bibitem{Single2016}
F.~Single, B.~Horstmann, A.~Latz, \emph{Phys. Chem. Chem. Phys.} \textbf{2016},
  \emph{18}, 17810.

\bibitem{Hein2020}
S.~Hein, T.~Danner, A.~Latz, \emph{ACS Appl. Energy Mater.} \textbf{2020},
  \emph{3}, 8519.

\bibitem{Weppner1977}
W.~Weppner, R.~A. Huggins, \emph{J. Electrochem. Soc.} \textbf{1977},
  \emph{124}, 1569.

\bibitem{Heubner2016}
C.~Heubner, M.~Schneider, A.~Michaelis, \emph{J. Electroanal. Chem.}
  \textbf{2016}, \emph{767}, 18.

\bibitem{Meddings2020}
N.~Meddings, M.~Heinrich, F.~Overney, J.~S. Lee, V.~Ruiz, E.~Napolitano,
  S.~Seitz, G.~Hinds, R.~Raccichini, M.~Gaberšček, J.~Park, \emph{J. Power
  Sources} \textbf{2020}, \emph{480}, 228742.

\bibitem{Sethurajan2015}
A.~K. Sethurajan, S.~A. Krachkovskiy, I.~C. Halalay, G.~R. Goward, B.~Protas,
  \emph{J. Phys. Chem. B} \textbf{2015}, \emph{119}, 12238.

\bibitem{Gouverneur2015}
M.~Gouverneur, J.~Kopp, L.~V. Wüllen, M.~Schönhoff, \emph{Phys. Chem. Chem.
  Phys.} \textbf{2015}, \emph{17}, 30680.

\bibitem{Horner2021}
J.~S. Horner, G.~Whang, D.~S. Ashby, I.~V. Kolesnichenko, T.~N. Lambert, B.~S.
  Dunn, A.~A. Talin, S.~A. Roberts, \emph{ACS Appl. Energy Mater.}
  \textbf{2021}, \emph{4}, 11460.

\bibitem{Bizeray2019}
A.~M. Bizeray, J.~H. Kim, S.~R. Duncan, D.~A. Howey, \emph{IEEE Trans. Control
  Syst. Technol.} \textbf{2019}, \emph{27}, 1862.

\bibitem{Bolay2021}
L.~J. Bolay, T.~Schmitt, S.~Hein, O.~S. Mendoza-Hernandez, E.~Hosono,
  D.~Asakura, K.~Kinoshita, H.~Matsuda, M.~Umeda, Y.~Sone, A.~Latz,
  B.~Horstmann, \emph{J. Power Sources Adv.} \textbf{2022}, \emph{14}, 100083.

\bibitem{Forman2012}
J.~C. Forman, S.~J. Moura, J.~L. Stein, H.~K. Fathy, \emph{J. Power Sources}
  \textbf{2012}, \emph{210}, 263.

\bibitem{Li2022}
W.~Li, I.~Demir, D.~Cao, D.~Jöst, F.~Ringbeck, M.~Junker, D.~U. Sauer,
  \emph{Energy Storage Mater.} \textbf{2022}, \emph{44}, 557.

\bibitem{Plett2006}
G.~L. Plett, \emph{J. Power Sources} \textbf{2006}, \emph{161}, 1369.

\bibitem{Plett2004}
G.~L. Plett, \emph{J. Power Sources} \textbf{2004}, \emph{134}, 277.

\bibitem{Aitio2020}
A.~Aitio, S.~G. Marquis, P.~Ascencio, D.~Howey, \emph{IFAC-PapersOnLine}
  \textbf{2020}, \emph{53}, 12497.

\bibitem{Sethurajan2019}
A.~Sethurajan, S.~Krachkovskiy, G.~Goward, B.~Protas, \emph{J. Comput. Chem.}
  \textbf{2019}, \emph{40}, 740.

\bibitem{Xu2022}
L.~Xu, X.~Lin, Y.~Xie, X.~Hu, \emph{Energy Storage Mater.} \textbf{2022},
  \emph{45}, 952.

\bibitem{Yamanaka2020}
T.~Yamanaka, Y.~Takagishi, T.~Yamaue, \emph{J. Electrochem. Soc.}
  \textbf{2020}, \emph{167}, 100516.

\bibitem{Bills2020}
A.~Bills, S.~Sripad, W.~L. Fredericks, M.~Guttenberg, D.~Charles, E.~Frank,
  V.~Viswanathan, \emph{ChemRxiv} \textbf{2020}.

\bibitem{Houchins2020}
G.~Houchins, V.~Viswanathan, \emph{J. Chem. Phys.} \textbf{2020}, \emph{153}.

\bibitem{Li2021}
W.~Li, J.~Zhang, F.~Ringbeck, D.~Jöst, L.~Zhang, Z.~Wei, D.~U. Sauer, \emph{J.
  Power Sources} \textbf{2021}, \emph{506}, 230034.

\bibitem{Zhao2020}
H.~Zhao, R.~D. Braatz, M.~Z. Bazant, \emph{J. Comput. Phys.} \textbf{2021},
  \emph{436}.

\bibitem{Krygier2021}
M.~C. Krygier, T.~LaBonte, C.~Martinez, C.~Norris, K.~Sharma, L.~N. Collins,
  P.~P. Mukherjee, S.~A. Roberts, \emph{Nat. Commun.} \textbf{2021}, \emph{12},
  1.

\bibitem{Biebl2019}
F.~Biebl, T.~Cvjetkovic, J.~oliver Schwarz, E.~Glatt, A.~Grießer, C.~Wagner,
  A.~Wiegmann, \emph{Conf. Ind. Comput. Tomogr.} \textbf{2019}, pages 1--4.

\bibitem{Furat2021}
O.~Furat, D.~P. Finegan, D.~Diercks, F.~Usseglio-Viretta, K.~Smith, V.~Schmidt,
  \emph{J. Power Sources} \textbf{2021}, \emph{483}, 229148.

\bibitem{Scharf2021}
J.~Scharf, M.~Chouchane, D.~P. Finegan, B.~Lu, C.~Redquest, M.~cheol Kim,
  W.~Yao, A.~A. Franco, D.~Gostovic, Z.~Liu, M.~Riccio, F.~Zelenka, J.-M. Doux,
  Y.~S. Meng, \emph{Nat. Nanotechnol.} \textbf{2022}, \emph{17}, 446.

\bibitem{Birkl2017}
C.~R. Birkl, M.~R. Roberts, E.~McTurk, P.~G. Bruce, D.~A. Howey, \emph{J. Power
  Sources} \textbf{2017}, \emph{341}, 373.

\bibitem{Tang2019}
S.~Tang, Z.~Wang, H.~Guo, J.~Wang, X.~Li, G.~Yan, \emph{Solid State Ionics}
  \textbf{2019}, \emph{343}, 115083.

\bibitem{Chen2020}
C.-H. Chen, F.~B. Planella, K.~O’Regan, D.~Gastol, W.~D. Widanage,
  E.~Kendrick, \emph{J. Electrochem. Soc.} \textbf{2020}, \emph{167}, 080534.

\bibitem{Maheshwari2016}
A.~Maheshwari, M.~A. Dumitrescu, M.~Destro, M.~Santarelli, \emph{J. Power
  Sources} \textbf{2016}, \emph{307}, 160.

\bibitem{Escalante2021}
J.~M. Escalante, S.~Sahu, J.~M. Foster, B.~Protas, \emph{Electrochem. Soc.}
  \textbf{2021}, \emph{168}, 110519.

\bibitem{Park2018}
S.~Park, D.~Kato, Z.~Gima, R.~Klein, S.~Moura, \emph{J. Electrochem. Soc.}
  \textbf{2018}, \emph{165}, A1309.

\bibitem{Zhao2020a}
J.~Zhao, M.~Cano, J.~J. Giner-Casares, R.~Luque, G.~Xu, \emph{Energy Environ.
  Sci.} \textbf{2020}, \emph{13}, 2618.

\bibitem{Gasper2021}
P.~Gasper, K.~Gering, E.~Dufek, K.~Smith, \emph{J. Electrochem. Soc.}
  \textbf{2021}, \emph{168}, 020502.

\bibitem{Barthelme2015}
S.~Barthelmé, N.~Chopin, V.~Cottet, \emph{Handbook of Approximate Bayesian
  Computation}, Chapman and Hall/CRC, 1st editio edition \textbf{2018}.

\bibitem{Gutmann2016}
M.~U. Gutmann, J.~Corander, \emph{J. Mach. Learn. Res.} \textbf{2016},
  \emph{17}, 1.

\bibitem{Schmitt2020}
T.~Schmitt, A.~Latz, B.~Horstmann, \emph{Electrochim. Acta} \textbf{2020},
  \emph{333}, 135491.

\bibitem{Marquis2019}
S.~G. Marquis, V.~Sulzer, R.~Timms, C.~P. Please, S.~J. Chapman, \emph{J.
  Electrochem. Soc.} \textbf{2019}, \emph{166}, A3693.

\bibitem{Doyle1993}
M.~Doyle, T.~F. Fuller, J.~Newman, \emph{J. Electrochem. Soc.} \textbf{1993},
  \emph{140}, 1526.

\bibitem{VanOijen2020}
M.~van Oijen, \emph{Bayesian Compendium}, Springer International Publishing,
  1st ed. 20 edition \textbf{2020}.

\bibitem{Held2010}
A.~Gelman, J.~Hill, \emph{Likelihood and Bayesian inference and computation},
  Springer Berlin Heidelberg, 2nd ed. 20 edition \textbf{2010}.

\bibitem{Subair2020}
S.~Subair, C.~Thron, \emph{Implementations and Applications of Machine
  Learning}, volume 782, Springer International Publishing, 1st ed. 20 edition
  \textbf{2020}.

\bibitem{Stone1974}
M.~Stone, \emph{J. R. Stat. Soc. B} \textbf{1974}, \emph{36}, 111.

\bibitem{Sunnaker2013}
M.~Sunnåker, A.~G. Busetto, E.~Numminen, J.~Corander, M.~Foll, C.~Dessimoz,
  \emph{PLoS Comput. Biol.} \textbf{2013}, \emph{9}.

\bibitem{Minka2001}
T.~P. Minka, \emph{MIT} \textbf{2001}, pages 1--75.

\bibitem{Minka2013}
T.~P. Minka, \emph{Proc. Conf. Uncertainty AI, 17th} \textbf{2013}, pages
  362--369.

\bibitem{Courbariaux2015}
M.~Courbariaux, Y.~Bengio, J.~P. David, \emph{Adv. Neural Inf. Process. Syst.}
  \textbf{2015}, \emph{2015-Janua}, 3123.

\bibitem{Nam2014}
W.~Nam, P.~Dollár, J.~H. Han, \emph{Adv. Neural Inf. Process. Syst.}
  \textbf{2014}, \emph{1}, 424.

\bibitem{Dahl2010}
G.~E. Dahl, M.~Ranzato, A.~R. Mohamed, G.~Hinton, \emph{Adv. Neural Inf.
  Process. Syst., 24th} \textbf{2010}, pages 1--9.

\bibitem{Loeliger2007}
H.~A. Loeliger, J.~Dauwels, J.~Hu, S.~Korl, L.~Ping, F.~R. Kschischang,
  \emph{Proc. IEEE} \textbf{2007}, \emph{95}, 1295.

\bibitem{Bilmes1998}
J.~Bilmes, \emph{Univ. Calif., Berkeley} \textbf{1998}, pages 1--281.

\bibitem{Kullback1951}
S.~Kullback, R.~A. Leibler, \emph{Ann. Math. Stat.} \textbf{1951}, \emph{22},
  79.

\bibitem{Alsing2019}
J.~Alsing, T.~Charnock, S.~Feeney, B.~Wandelt, \emph{Mon. Not. R. Astron. Soc.}
  \textbf{2019}, \emph{488}, 4440.

\bibitem{Fer2018}
I.~Fer, R.~Kelly, P.~R. Moorcroft, A.~D. Richardson, E.~M. Cowdery, M.~C.
  Dietze, \emph{Biogeosciences} \textbf{2018}, \emph{15}, 5801.

\bibitem{Arnold2017}
B.~J. Arnold, M.~U. Gutmann, Y.~H. Grad, S.~K. Sheppard, J.~Corander,
  M.~Lipsitch, W.~P. Hanage, \emph{bioRxiv, Genet.} \textbf{2018}, \emph{208},
  1247.

\bibitem{Kangasraasio2017}
A.~Kangasrääsiö, K.~Athukorala, A.~Howes, J.~Corander, S.~Kaski,
  A.~Oulasvirta, \emph{Conf. Hum. Factors Comput. Syst. Proc.} \textbf{2017},
  \emph{2017-May}, 1295.

\bibitem{Lintusaari2018}
J.~Lintusaari, H.~Vuollekoski, A.~Kangasrääsiö, K.~Skytén, M.~Järvenpää,
  P.~Marttinen, M.~U. Gutmann, A.~Vehtari, J.~Corander, S.~Kaski, \emph{J.
  Mach. Learn. Res.} \textbf{2018}, \emph{19}, 1.

\bibitem{Lintusaari2017}
J.~Lintusaari, M.~U. Gutmann, R.~Dutta, S.~Kaski, J.~Corander, \emph{Syst.
  Biol.} \textbf{2017}, \emph{66}, e66.

\bibitem{Sobol1967}
I.~M. Sobol', \emph{USSR Comput. Math. Math. Phys.} \textbf{1967}, \emph{7},
  86.

\bibitem{Danner2016}
T.~Danner, M.~Singh, S.~Hein, J.~Kaiser, H.~Hahn, A.~Latz, \emph{J. Power
  Sources} \textbf{2016}, \emph{334}, 191.

\bibitem{Birkl2015}
C.~R. Birkl, E.~McTurk, M.~R. Roberts, P.~G. Bruce, D.~A. Howey, \emph{J.
  Electrochem. Soc.} \textbf{2015}, \emph{162}, A2271.

\bibitem{Mistry2021}
A.~Mistry, S.~Trask, A.~Dunlop, G.~Jeka, B.~Polzin, P.~P. Mukherjee,
  V.~Srinivasan, \emph{J. Electrochem. Soc.} \textbf{2021}, \emph{168}, 070536.

\bibitem{Sun2021}
Y.-K. Sun, \emph{ACS Energy Lett.} \textbf{2021}, \emph{6}, 2187.

\bibitem{Patel2003}
K.~K. Patel, J.~M. Paulsen, J.~Desilvestro, \emph{J. Power Sources}
  \textbf{2003}, \emph{122}, 144.

\bibitem{Nyman2008}
A.~Nyman, M.~Behm, G.~Lindbergh, \emph{Electrochim. Acta} \textbf{2008},
  \emph{53}, 6356.

\bibitem{Huang2020}
S.~Huang, J.~M. Cole, \emph{Sci. Data} \textbf{2020}, \emph{7}, 1.

\bibitem{Landesfeind2019}
J.~Landesfeind, H.~A. Gasteiger, \emph{J. Electrochem. Soc.} \textbf{2019},
  \emph{166}, A3079.

\bibitem{Colclasure2010}
A.~M. Colclasure, R.~J. Kee, \emph{Electrochim. Acta} \textbf{2010}, \emph{55},
  8960.

\bibitem{Latz2013}
A.~Latz, J.~Zausch, \emph{Electrochim. Acta} \textbf{2013}, \emph{110}, 358.

\bibitem{Kurchin2020}
R.~Kurchin, V.~Viswanathan, \emph{J. Chem. Phys.} \textbf{2020}, \emph{153}.

\bibitem{Fraggedakis2021}
D.~Fraggedakis, M.~McEldrew, R.~B. Smith, Y.~Krishnan, Y.~Zhang, P.~Bai, W.~C.
  Chueh, Y.~Shao-Horn, M.~Z. Bazant, \emph{Electrochim. Acta} \textbf{2021},
  \emph{367}, 137432.

\bibitem{Xing2020}
F.~Xing, S.~Ma, L.~Zhang, \emph{J. Phys. Chem. C} \textbf{2020}, \emph{124},
  10832.

\bibitem{Chien2021}
Y.-C. Chien, H.~Liu, A.~S. Menon, W.~R. Brant, D.~Brandell, M.~J. Lacey,
  \emph{ChemRxiv} \textbf{2021}, pages 1--8.

\bibitem{Bergstrom2021}
H.~K. Bergstrom, K.~D. Fong, B.~D. McCloskey, \emph{J. Electrochem. Soc.}
  \textbf{2021}, \emph{168}, 060543.

\bibitem{Geng2022}
Z.~Geng, Y.~C. Chien, M.~J. Lacey, T.~Thiringer, D.~Brandell,
  \emph{Electrochim. Acta} \textbf{2022}, \emph{404}, 1.

\bibitem{Gelman2013}
J.~K. Kruschke, \emph{Bayesian data analysis}, volume~1, Taylor \& Francis,
  third edit edition \textbf{2010}.

\bibitem{Mistry2021a}
A.~Mistry, A.~Verma, S.~Sripad, R.~Ciez, V.~Sulzer, F.~B. Planella, R.~Timms,
  Y.~Zhang, R.~Kurchin, P.~Dechent, W.~Li, S.~Greenbank, Z.~Ahmad,
  D.~Krishnamurthy, A.~M.~F. Jr., K.~Tenny, P.~Patel, D.~J. Robles, P.~Gasper,
  A.~Colclasure, A.~Baskin, C.~D. Scown, V.~R. Subramanian, E.~Khoo, S.~Allu,
  D.~Howey, S.~DeCaluwe, S.~A. Roberts, V.~Viswanathan, \emph{ACS Energy Lett.}
  \textbf{2021}, \emph{6}, 3831.

\bibitem{Wang2002}
Z.~J. Wang, \emph{J. Comput. Phys.} \textbf{2002}, \emph{178}, 210.

\bibitem{Andersson2019}
J.~A. Andersson, J.~Gillis, G.~Horn, J.~B. Rawlings, M.~Diehl, \emph{Math.
  Program. Comput.} \textbf{2019}, \emph{11}, 1.

\bibitem{Sulzer2020}
V.~Sulzer, S.~G. Marquis, R.~Timms, M.~Robinson, S.~J. Chapman, \emph{J. Open
  Res. Software} \textbf{2021}, \emph{9}, 1.

\bibitem{Hoffman2014}
M.~D. Hoffman, A.~Gelman, \emph{J. Mach. Learn. Res.} \textbf{2014}, \emph{15},
  1593.

\bibitem{Schmalstieg2018}
J.~Schmalstieg, C.~Rahe, M.~Ecker, D.~U. Sauer, \emph{J. Electrochem. Soc.}
  \textbf{2018}, \emph{165}, A3799.

\bibitem{Wang2021}
A.~A. Wang, S.~E.~J. O'Kane, F.~B. Planella, J.~L. Houx, K.~O'Regan, M.~Zyskin,
  J.~Edge, C.~W. Monroe, S.~J. Cooper, D.~A. Howey, E.~Kendrick, J.~M. Foster,
  \emph{Prog. Energy} \textbf{2022}, \emph{4}, 032004.

\bibitem{Kang2021}
S.~D. Kang, W.~C. Chueh, \emph{J. Electrochem. Soc.} \textbf{2021}, \emph{168},
  120504.

\bibitem{Lai2021}
Q.~Lai, H.~J. Ahn, Y.~J. Kim, Y.~N. Kim, X.~Lin, \emph{Appl. Energy}
  \textbf{2021}, \emph{295}, 117034.

\bibitem{Less2012}
G.~B. Less, J.~H. Seo, S.~Han, A.~M. Sastry, J.~Zausch, A.~Latz, S.~Schmidt,
  C.~Wieser, D.~Kehrwald, S.~Fell, \emph{J. Electrochem. Soc.} \textbf{2012},
  \emph{159}, A697.

\bibitem{Kirk2021}
T.~L. Kirk, C.~P. Please, S.~J. Chapman, \emph{J. Electrochem. Soc.}
  \textbf{2021}, \emph{168}, 060554.

\bibitem{Single2019}
F.~Single, B.~Horstmann, A.~Latz, \emph{J. Phys. Chem. C} \textbf{2019},
  \emph{123}, 27327.

\bibitem{Luck2019}
J.~Lück, A.~Latz, \emph{Phys. Chem. Chem. Phys.} \textbf{2019}, \emph{21},
  14753.

\bibitem{Ikonomov2020}
B.~Ikonomov, M.~U. Gutmann, \emph{Proc. Mach. Learn. Res.} \textbf{2019},
  \emph{108}, 2819.

\end{thebibliography}

\clearpage

%%%%%%%		TOC Entry			%%%%%%% 

\section*{Entry for the Table of Contents}

%	 please select one option only and delete the other one

%%%%%%%		Option 1			%%%%%%%    

\noindent\rule{11cm}{2pt}
\begin{minipage}{5.5cm}
\includegraphics[width=5.5cm]{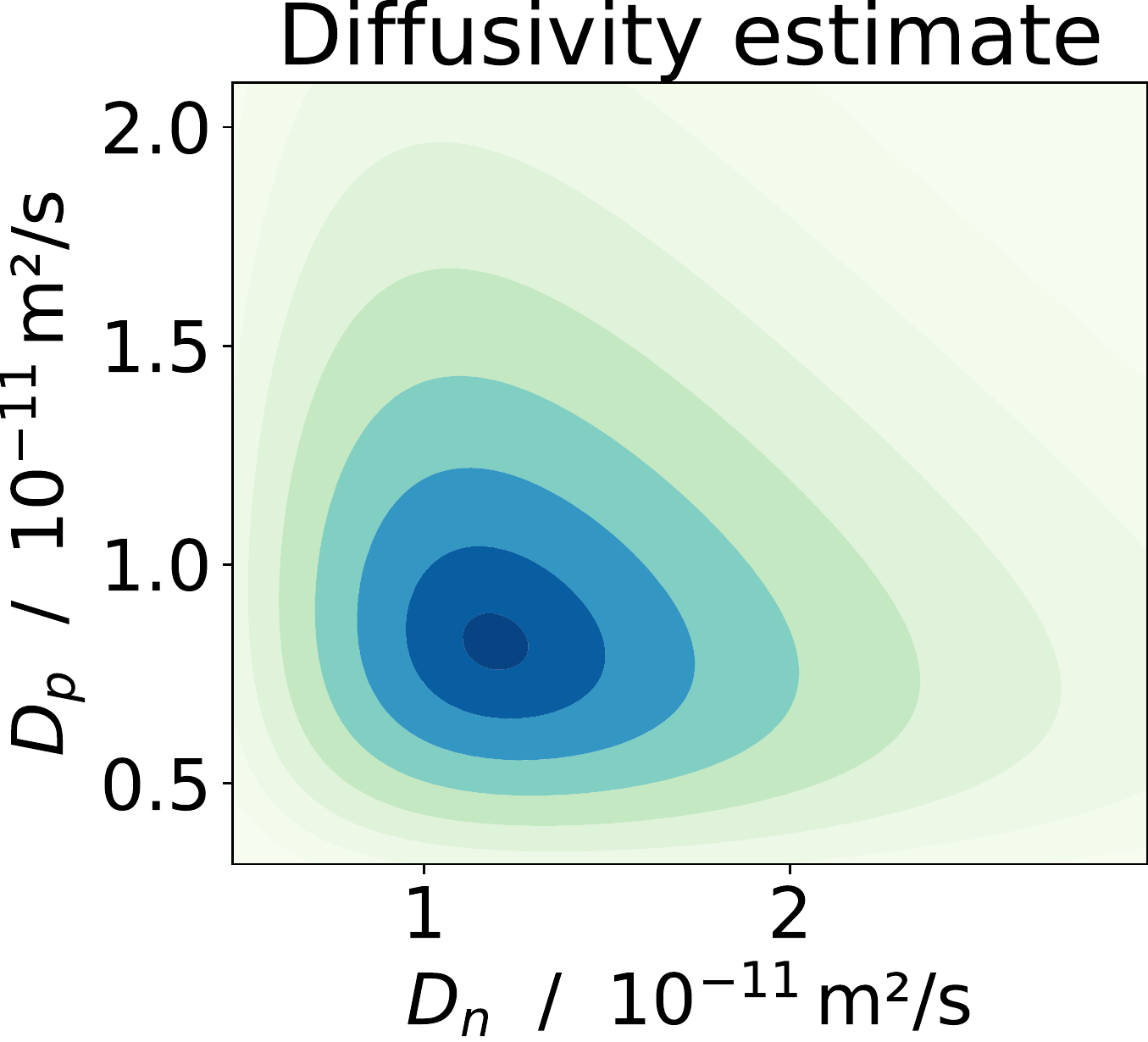} 
\end{minipage}
%\hspace{0.5cm}
\begin{minipage}{5.5cm}
\large\textsf{We introduce a computer algorithm that incorporates the experience of battery researchers to extract information from experimental data reproducibly. This enables the fitting of complex models that take up to a few minutes to simulate. For validation, we process full-cell GITT measurements to characterize the diffusivities of both electrodes non-destructively.}
\end{minipage}
\noindent\rule{11cm}{2pt}

\vspace{2cm}

%%%%%%%%%%%%%%%%%%%%%%%%%%%%%%%%%%%%%%%%%%%%%%%%%%%%%%%%%%
%%%%%%%%%%%%%%%%%%%%%%%%%%%%%%%%%%%%%%%%%%%%%%%%%%%%%%%%%%
%%%%%%%%%%%%%%%%%%%%%%%%%%%%%%%%%%%%%%%%%%%%%%%%%%%%%%%%%%

\end{document}